\documentclass[aps,prd,superscriptaddress,floatfix,nofootinbib]{revtex4}

\usepackage{verbatim}
\usepackage{mathtools}
\usepackage{ragged2e}
\usepackage{amsmath}
\usepackage{amsfonts}
\usepackage{amssymb}
\usepackage{graphicx}
\usepackage{slashed}
\usepackage{bm}
\usepackage{color}
\usepackage{epsf}
\usepackage{orcidlink}
\usepackage[parfill]{parskip}
\usepackage{nicefrac}
\usepackage{booktabs}
\usepackage{multirow}
\usepackage[normalem]{ulem}
\usepackage{mathrsfs}

\usepackage{hyperref}
\hypersetup{
  colorlinks   = true, 
  urlcolor     = blue, 
  linkcolor    = black, 
  citecolor   = red 
}

\newcommand{\quark}{\mathfrak{q}}

\newcommand{\eps}{\varepsilon} 
\newcommand{\M}{\mathcal{M}} 
\newcommand{\amp}{\mathcal{A}} 

\newcommand{\wpbbiii}{\widetilde{\mathbb{P}}_{\rm (iii)}}

\newcommand{\bq}{\bar{q}}
\newcommand{\bp}{\bar{p}}

\newcommand{\scale}{\mathbb{Q}}

\newcommand{\dxi}{\partial_{\xi}}

\newcommand\Li[2]{{\,\textrm{Li}_{#1}\left(#2\right)}}  
\newcommand\Ln[1]{{\,\textrm{ln}\left(#1\right)}}  

\usepackage{tcolorbox}
\tcbuselibrary{listings, breakable, skins}
\newtcolorbox{eqbox}[2][]{%
	enhanced,
	colback=blue!5!white,
	colframe=blue!75!black,
	fonttitle=\bfseries,
	title=#2,
	sharp corners=south,
	boxrule=0.8pt,
	drop shadow,
	breakable,
	#1
}

\DeclareUnicodeCharacter{2212}{-}

\usepackage{cancel}

\begin{document}

\title{Coherent deeply virtual Compton scattering on helium-4 beyond leading power}

\author{V.~Mart\'inez-Fern\'andez\,\orcidlink{0000-0002-0581-7154}}
\affiliation{IRFU, CEA, Universit\'e Paris-Saclay, F-91191 Gif-sur-Yvette, France}
\affiliation{Center for Frontiers in Nuclear Science, Stony Brook University, Stony Brook, NY 11794, USA}
\author{B.~Pire\,\orcidlink{0000-0003-4882-7800}}
\affiliation{Centre de Physique Th\'eorique, CNRS, École polytechnique, I.P. Paris, 91128 Palaiseau, France  }

\author{P.~Sznajder\,\orcidlink{0000-0002-2684-803X}}
\affiliation{National Centre for Nuclear Research (NCBJ), Pasteura 7, 02-093 Warsaw, Poland}

\author{J.~Wagner\,\orcidlink{0000-0001-8335-7096}}
\affiliation{National Centre for Nuclear Research (NCBJ), Pasteura 7, 02-093 Warsaw, Poland}


\begin{abstract}

Coherent hard exclusive reactions on light nuclei provide access to their quark and gluon structure and enable three-dimensional tomography of these complex systems. We study deeply virtual Compton scattering on a helium-4 target, including both kinematic twist-3 and twist-4 corrections, as well as next-to-leading-order corrections to the twist-2 amplitude in the strong coupling $\alpha_s$. We show that these contributions are crucial for achieving a precise description of the data and, as a result, obtain the first tomographic image of the helium-4 nucleus at the quark-gluon level.

\end{abstract}

\maketitle

\section{Introduction}\label{sec::intro}

\begin{figure}
	\includegraphics[scale=1]{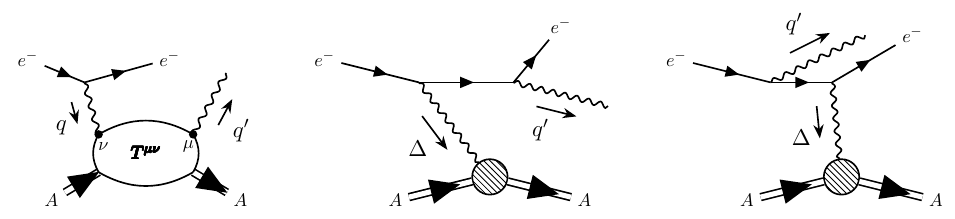}
	\caption{Deeply virtual Compton scattering (DVCS, on the left) and the two cases of Bethe-Heitler background, BH (center) and BHX (right). The tensor $T^{\mu\nu}$ is the so-called Compton tensor, see main text.}
	\label{fig::subprocesses}
\end{figure}

Studying coherent exclusive scattering has long  been recognized~\cite{Berger:2001zb, Kirchner:2003wt, Guzey:2003jh,Cano:2003ju, Scopetta:2004kj} as the best way to progress in our understanding of the quark and gluon structure of light nuclei, and therefore as a key step toward the understanding of the confinement mechanism beyond the nucleon or meson case. The golden channel, as in the nucleon case, is
the {\it electroproduction of a real photon}~\cite{Belitsky:2001ns}
\begin{equation}\label{reaction}
	e^-(k) + A(p)\to e^-(k') + A(p') + \gamma(q')\,,\quad q'^2=0\,,
\end{equation}
where  $e^-$ stands for the electron, $A$ represents the nucleus, and $\gamma$ is the produced real photon, while the momenta of the corresponding particles are given in parentheses. This process can be understood as the superposition of two types of scatterings depicted in Fig.~\ref{fig::subprocesses}: a purely electromagnetic {\it Bethe-Heitler} process where the final-state photon is the result of leptonic bremsstrahlung and it is characterized by the electromagnetic form factors (EFFs) of the target; and the {\it deeply virtual Compton scattering} (DVCS) consisting of an absorption of a virtual photon and subsequent emission of a real one. 

The DVCS amplitude is conveniently described in terms of {\it Compton Form Factors} (CFFs), which are complex-valued convolutions of perturbatively calculable coefficient functions, describing the hard parton--photon interaction, with generalized parton distributions (GPDs)~\cite{PhysRevLett.78.610,Belitsky:2005qn,Diehl:2003ny}.
This factorized form of the scattering amplitude applies in the generalized Bjorken regime, where the momentum transfer to the target remains small in comparison with the photon virtuality. Denoting the latter by $-q^2 = -(k-k')^2 = Q^2>0$ and the former by the Mandelstam variable $t = (p'-p)^2 < 0$, one finds that the leading contribution, referred to as the leading-twist (LT) approximation, is obtained in the limit $|t|/Q^2 \to 0$.

 The real experiments obviously do not realize the leading-twist (LT) limit. Consequently, deviations from this limit manifest as corrections that enter the coefficient functions of DVCS in powers of the form $(|t|/Q^2)^{(\tau-2)/2}$, commonly referred to as {\it kinematic-twist} ($\tau$) {\it corrections}~\cite{Braun:2025xlp,Martinez-Fernandez:2025gub}.\footnote{Although they are sometimes called target mass corrections, we explicitly show below that their effect is not proportional to the nucleus mass, but rather to the momentum transfer.} The importance of systematically accounting for these corrections is multifaceted. Including them improves theoretical accuracy and enables the use of a larger portion of experimental data for the extraction of CFFs, thereby relaxing kinematic cuts that remove points affected by higher-twist contributions. Besides these kinematic corrections, there are also genuine higher-twist contributions associated with matrix elements of higher-twist operators. These are beyond the scope of the present study, in which we focus exclusively on the kinematic-twist sector.

Furthermore, {\it hadron and nuclear tomography}~\cite{Burkardt:2002hr,Ralston:2001xs,Diehl:2002he}, which provides transverse imaging of their internal structure, requires integration over all possible values of $t$. The inclusion of kinematic $t$-dependent corrections is therefore essential for a consistent and complete tomographic analysis. Similarly, since GPDs depend explicitly on the momentum transfer $t$, incorporating recoil effects in data analyses is crucial for achieving a more reliable modeling of these distributions, done thus far under a leading twist formalism.

In addition, moments of GPDs are related to the form factors parameterizing the QCD energy-momentum tensor, known as gravitational form factors, which provide access to the mechanical properties of hadrons and nuclei~\cite{Lorce:2018egm}. Recent works~\cite{Martinez-Fernandez:2025rcg,Martinez-Fernandez:2025jvk} extended the relation between Compton form factors and gravitational form factors beyond leading twist by incorporating kinematic-twist corrections, thus allowing Compton scattering observables to probe a broader class of gravitational form factors.

In summary, twist effects are not only relevant for precision analyses, but they also provide unique insights into the mechanical structure of hadronic matter. As a result, recovering these recoil and mass effects must become an integral part of the standard for theoretical calculations, alongside with next-to-leading order (NLO) corrections in the strong coupling constant $\alpha_s$ whose relevance have already been highlighted~\cite{Pire:2011st}. Such corrections also include contributions of the gluon-transversity GPD responsible for a two-unit photon-helicity flip at LT~\cite{Belitsky:2000jk}, making it a competing effect to the LO twist-4 contribution by quarks to said helicity flip.

This motivates revisiting DVCS with modern theoretical tools, combining higher-twist and NLO accuracy with phenomenological studies. As a first step, in this manuscript and in the companion letter~\cite{Martinez-Fernandez:2026} we focus on DVCS off a spinless target, and in particular on helium-4. Compared to the nucleon, spinless targets are described by a reduced set of  GPDs, which makes them a natural starting point for such an analysis. Among them, helium-4 is especially interesting, as it provides a clean setting to explore nuclear effects in hadron structure. Moreover, existing JLab measurements~\cite{PhysRevLett.119.202004,PhysRevC.104.025203} and ongoing experimental activity make this study both timely and phenomenologically relevant.

Our line of reasoning for building a model of helium-4 GPDs is very different from the ones used in \cite{Fucini_2018} where the nucleus wave function is convoluted with nucleon GPDs. In contrast to this approach, ours respects Lorentz invariance and treats the nuclear state as a QCD object without any prejudice on its nucleon decomposition. A comparison of the tomographic picture of the nucleus based on the same experimental data but with these different approaches would be very welcome, but is outside the scope of our study

This manuscript is organized as follows. Sect.~\ref{sec::crossSection} presents the calculation of the cross section and the corresponding amplitudes at twist-4 and NLO accuracy. Sect.~\ref{sec::4Hemodel} is devoted to the modeling of helium-4 GPDs. The results of the fit to DVCS data are presented in Sect.~\ref{sec::Results}, including the first tomographic image of the helium-4 nucleus in terms of quarks and gluons, as well as numerical estimates of cross sections and amplitudes for JLab experiments. Sect.~\ref{sec::conclusions} concludes with a summary of our main results and final remarks. For completeness, App.~\ref{sec::kinematics} collects the relevant kinematics,  App.~\ref{app:BH} provides details of the amplitude calculations based on the Kleiss--Stirling spinor techniques~\cite{KLEISS198461,KLEISS1985235} that we use in Sect.~\ref{sec::crossSection}, and we collect in App.~\ref{app:CFs} the expressions for helicity amplitudes.
\section{Amplitudes and cross-section}\label{sec::crossSection}

\begin{figure}
	\centering
	\includegraphics[scale=1]{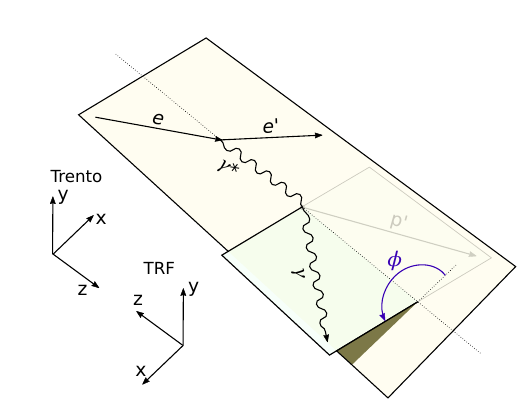}
	\caption{Target rest frame (TRF) and Trento frame.}
	\label{fig::TRFandTrento}
\end{figure}

The differential cross section for a beam of charge sign $\chi$ and helicity $s$, producing a real photon with polarization $\lambda$, can be written as: 
\begin{equation}\label{cross_section}
	\frac{d^{4}\sigma^{s\lambda}_\chi}{d{x_A}~dQ^2~dt~d\phi}
	= \frac{\alpha_{\rm EM}^3}{8\pi} \frac{x_A y^2}{Q^4\sqrt{1+\omega^2}}
	\left|\frac{\mathcal{M}^{s\lambda}_\chi}{e^3}\right|^2 \,,
\end{equation}
where the inelasticity variable is given by $y=pq/(pk)$, $\omega = 2x_A M/Q$, $M$ is the nucleus mass, the nuclear Bjorken variable $x_A$ is defined as
\begin{equation}\label{xA_vs_xB}
	x_A = \frac{Q^2}{2pq} \approx \frac{x_B}{A}\,,
\end{equation}
and the approximate relation to the proton Bjorken variable $x_B$ holds for a nucleus with mass number $A$ (up to binding energy corrections). The angle $\phi$ is defined according to the Trento convention~\cite{Bacchetta:2004jz} as presented in Fig.~\ref{fig::TRFandTrento}. The detailed parametrization of all four-momenta in the target rest frame (TRF) of Ref.~\cite{Belitsky:2001ns}, which is related to the standard Trento frame by a $180^\circ$ rotation about the $y$-axis, is briefly recalled in App.~\ref{sec::kinematics}. 

The total amplitude is given by the sum of the three contributions corresponding to the diagrams shown in Fig.~\ref{fig::subprocesses}:
\begin{equation}\label{total_iM}
	i\M^{s\lambda}_{\chi}= i\M^{s\lambda}_{\chi,\,\rm DVCS}+ i\M^{s\lambda}_{\rm BH} + i\M^{s\lambda}_{\rm BHX}\,.
\end{equation}
In the following, we describe the calculation of each of these contributions.
\subsection{DVCS amplitude}
The amplitude for the DVCS process, shown in the left panel of Fig.~\ref{fig::subprocesses}, is given by: 
\begin{equation}
	i\M^{s\lambda}_{\chi,\,\rm DVCS} = \frac{i e^3 \chi}{q^2}\bar{u}(k',s)\gamma_\nu u(k,s) T^{\mu\nu}\eps'_\mu(-\lambda)\,,
\end{equation}
where the Compton tensor $T^{\mu\nu}$ ($\bq=(q+q')/2$ and $\mathcal{T}$ represents the time ordering) 
\begin{align}
	T^{\mu\nu} & = i\int d^4 z\ e^{i \bar{q} z} \langle p'| \mathcal{T}\left\{ j^\mu(z/2)j^\nu(-z/2) \right\} |p\rangle \nonumber\\
	& = i\int d^4z\ e^{i q' z} \langle p'| \mathcal{T}\left\{ j^\mu(z)j^\nu(0) \right\} |p\rangle \nonumber\\
	& = \sum_{\Lambda,\Lambda'} T_{(\Lambda\Lambda')}^{\mu\nu}\amp^{\Lambda\Lambda'}\,,
\end{align}
can be parameterized by means of photon-helicity-dependent amplitudes $\amp^{\Lambda\Lambda'}$ ~\cite{Braun:2012bg,Martinez-Fernandez:2025gub}. For an incoming photon with polarization $\Lambda$ and an outgoing one with polarization $\Lambda'$, these take the form
\begin{equation}\label{A^AB}
	\amp^{\Lambda\Lambda'} = (-1)^{\Lambda-1} \left( \eps'_\mu(\Lambda') \right)^* T^{\mu\nu} \eps_\nu(\Lambda) \,,
\end{equation}
where the polarization vectors are defined in Eqs.~(\ref{epsPrime}) and (\ref{eps0}). After accounting for parity symmetry, the sum over photon polarizations is limited to $\Lambda\Lambda'\in\{++,+-,0+\}$, where $\pm$ stand for transverse polarizations while $0$ for the longitudinal one. The corresponding Lorentz structures, expressed by means of the polarization vectors 
\begin{align}
	T_{(++)}^{\mu\nu} & = \eps^\mu(+)\eps^\nu(-) + \eps^\mu(-)\eps^\nu(+) \,, \label{T++munu_eps} \\
	T_{(+-)}^{\mu\nu} & = \eps^\mu(+)\eps^\nu(+) + \eps^\mu(-)\eps^\nu(-) \,, \label{T+-munu_eps} \\
	T_{(0+)}^{\mu\nu} & = \eps^\mu(+)\eps^\nu(0) + \eps^\mu(-)\eps^\nu(0)  \label{T0+munu_eps} \,,
\end{align}
take the form
\begin{align}
	T_{(++)}^{\mu\nu} & = -g_\perp^{\mu\nu} = -\left( g^{\mu\nu} - \frac{n^\mu n'^\nu + n^\nu n'^\mu}{nn'} \right)\,, \label{T++munu} \\
	T_{(+-)}^{\mu\nu} & = \frac{1}{|\bp_\perp|^2}\left[ p_\perp^\mu p_\perp^\nu - \widetilde{p}_\perp^\mu \widetilde{p}_\perp^\nu \right] \,, \label{T+-munu} \\
	T_{(0+)}^{\mu\nu} & = \frac{\sqrt{2}}{|\bp_\perp|}p_\perp^\mu\left[ \frac{1}{Q}q^\nu - \frac{2Q}{\scale^2}q'^\nu \right]\,, \label{T0+munu}
\end{align}
where our choice of longitudinal vectors $n$ and $n'$ is described in App.~\ref{sec::kinematics}, as well as the exact parametrization of the transverse momenta $p_\perp$ and $\widetilde{p}_\perp$. 
We also introduced the hard scale of the process given by $\scale^2=Q^2+t$ and the dual transverse four-vector $\widetilde{p}_\perp^\mu = \epsilon^{\mu\nu}_\perp p_\nu$ by means of the transverse Levi-Civita tensor: $\epsilon_\perp^{\mu\nu} = \epsilon^{\mu\nu\alpha\beta}n_\alpha n'_\beta/(nn')$.

We can express our final result for the DVCS amplitude in the following way:
\begin{align}\label{iM_DVCS}
	i\M_{\chi,\,\rm DVCS}^{s\lambda}= &\ \frac{-ie^3\chi}{Q^2} \Bigg[ \amp^{++} \left(\frac{\sqrt{2}}{2|\bp_\perp|}\right) \left( \sum_{j=1}^3 B_j g(s,k',L_j,k) - i\lambda p_\mu\epsilon_\perp^{\mu\nu} j_\nu(s,k',k) \right) \nonumber\\
	&\phantom{\ \frac{ie^3\chi}{Q^2-i0} \Bigg[ } + \amp^{+-}\left(\frac{\sqrt{2}}{2|\bp_\perp|}\right)  \left( \sum_{j=1}^3 B_j g(s,k',L_j,k) + i\lambda p_\mu\epsilon_\perp^{\mu\nu} j_\nu(s,k',k) \right) \nonumber\\ 
	& \phantom{\ \frac{ie^3\chi}{Q^2-i0} \Bigg[ }  - \amp^{0+} \frac{2Q}{\scale^2} g(s,k',q',k) \Bigg] \,,
\end{align}
where $B_j\in\{ 1,-p^-Q^2,-\alpha \}$, $L_j\in\{r_2,q',n'\}$, and the functions $g$ and $j_\mu$ are defined by Eqs.~(\ref{g}) and (\ref{j}), respectively. If the CFFs are kept to twist-4 accuracy, as it is the case of Eqs.~(\ref{A++})-(\ref{A0+}), then in the last line we can approximate $2Q/\scale^2 \approx 2/Q$ since $\scale^2=Q^2+t$. Otherwise, the expression~(\ref{iM_DVCS}) is exact to any order in kinematic twist and power of $\alpha_s$.

In our analysis, we take into account twist-3 and twist-4 corrections, as well as NLO corrections to the leading-twist contribution. We write them {\it schematically} as:
\begin{align}
	\amp^{++}(\xi,t)= &
    \ \int_{-1}^1 dx\ \frac{1}{\xi}\Bigg\{ \left[ 
        C_0(x/\xi) +\frac{\alpha_SC_F}{4\pi}~{C^q_{NLO}}(x/\xi)  +\frac{t}{\scale^2}~{C_{HT}}(x,\xi)
    \right]H^\Sigma(x,\xi,t) 
    \nonumber\\
      & 
      \phantom{AAAAAAAa}
      +\frac{\alpha_ST_F}{4\pi}\frac{1}{\xi}~{C^g_{NLO}}(x/\xi) ~H^g(x,\xi,t) \Bigg\}
	\,, \label{A++} \\
	\amp^{+-}(\xi,t) = & 
     \ \int_{-1}^1 dx\ \frac{1}{\xi}\Bigg\{ 
     \frac{t}{\scale^2}~ C^{+-}_{HT}(x,\xi)~H^\Sigma(x,\xi,t) 
     +\frac{\alpha_ST_F}{4\pi} \frac{1}{\xi}~C^{+-}_{NLO}(x/\xi)~H^g_T(x,\xi,t) \Bigg\} 
    \,, \label{A+-} \\
	\amp^{0+}(\xi,t) = &
     \ \int_{-1}^1 dx\ \frac{1}{\xi}\Bigg\{ 
     \frac{\sqrt{-t}}{\scale}~ C^{0+}_{HT}(x,\xi)~H^\Sigma(x,\xi,t)  \Bigg\} 
    \,, \label{A0+}
\end{align}
where
\begin{equation}
    H^\Sigma(x,\xi,t)= \sum_{q \in \{u,d,s\}} e_q^2\,H^q (x,\xi,t) \,.
\end{equation}
Our definitions of gluon $H^g$ and quark $H^q$  GPDs follow the review~\cite{Diehl:2003ny} with the skewness variable as in Eq.~(\ref{xi_exact}), and $e_q$ the electric charge of quark $q$ relative to the positron. The NLO coefficient functions, as well as those at the HT are known~\cite{Ji:1998xh, Pire:2011st,Braun:2012bg,Braun:2014sta,Martinez-Fernandez:2025gub,Belitsky:2000jk}, and were, for completeness, gathered and explained in detail in App.~\ref{app:CFs}.

\subsection{The  Bethe-Heitler amplitudes}
\label{sec::amplitudes_BH}

The detailed derivation of the expressions for BH amplitudes is presented in the App.~\ref{app:BH}. Here we shortly summarize the results. For the Feynman diagram on the middle of Fig.~\ref{fig::subprocesses}, the corresponding amplitude reads
\begin{align}\label{iM_BH}
	i\M^{s\lambda}_{\rm BH} =&\ \frac{i e^3 F(t)}{  (k'+q')^2   t  } \left(\frac{\sqrt{2}}{2|\bp_\perp|}\right) \nonumber\\
	&\ \times \sum_{i=1}^2 \sum_{\ell=1}^4 g(s,R_i,\widetilde{r}_\ell,k) \left[ \sum_{j=1}^3 B_j g(s,k',L_j,R_i) - i\lambda p_\nu\epsilon_\perp^{\nu\rho}j_\rho(s,k',R_i) \right] \,,
\end{align}
where $B_j\in\{ 1,-p^-Q^2,-\alpha \}$, $L_j\in\{r_2,q',n'\}$, $R_i\in\{k',q'\}$ and $\widetilde{r}_\ell\in\{r_1,r_2,r'_1,r'_2\}$.

Conversely, for the right plot in Fig.~\ref{fig::subprocesses},
\begin{align}\label{iM_BHX}
	i\M_{\rm BHX}^{s\lambda} =&\ \frac{i e^3 F(t)}{(k-q')^2 t } \left( \frac{\sqrt{2}}{2|\bp_\perp|} \right) \nonumber\\
	&\ \times \sum_{i=1}^2 \sum_{\ell=1}^4 \sigma(\widetilde{R}_i) g(s,k',\widetilde{r}_\ell,\widetilde{R}_i) \left[ \sum_{j=1}^3 B_j g(s,\widetilde{R}_i,L_j,k) - i\lambda p_\nu \epsilon_\perp^{\nu\rho}j_\rho(s,\widetilde{R}_i,k) \right] \,,
\end{align}
where, again $B_j\in\{ 1,-p^-Q^2,-\alpha \}$, $L_j\in\{r_2,q',n'\}$ and $\widetilde{r}_\ell\in\{r_1,r_2,r'_1,r'_2\}$, while $\widetilde{R}_i\in\{k,q'\}$ carry the signature function $\sigma(\widetilde{R}_i) \in\{+1,-1\}$, respectively.

\section{Model for helium-4 GPDs}\label{sec::4Hemodel}

Let us now model the helium-4 GPDs as rigorously as possible, while maintaining a sufficiently flexible parametrization so that it reproduces existing data, has the potential to describe future data, and allows for the reliable extraction of tomographic features. As in the case of modeling nucleon GPDs (see, for instance, Ref.~\cite{Goloskokov:2006hr}), the fundamental property of polynomiality is preserved by starting from double distributions. The limits $(\xi,t) \to (0,0)$ of GPDs $H^{q,g}$ are readily enforced by using a Radyushkin-type Ansatz~\cite{Radyushkin:1998es} with nuclear PDFs extracted from experimental data~\cite{AbdulKhalek:2022fyi}. For valence quarks, we require that the measured dependence on the $t$ variable of the electromagnetic form factor is recovered as the first $x$-moment of the GPD. For sea quarks and gluons, the dependence on $t$ will be constrained by sparse coherent DVCS experimental data. In all these cases, we rely on isospin symmetry, assuming an exact symmetry between up and down quarks. A somewhat different approach is adopted for modeling the GPD $H_{T}^{g}$. Given the limited empirical constraints currently available for this distribution, we primarily rely on theory-driven constraints in this case.

Our models for helium-4 GPDs, $H^{i}(x,\xi,t; \mu^2)$, where $i = \{u_{\mathrm{val}}, u_{\mathrm{sea}}, d_{\mathrm{val}}, d_{\mathrm{sea}}, s, g\}$ denotes the parton type explicitly distinguishing between valence and sea components, are based on double distributions, $F^{i}(\beta, \alpha, t; \mu^2)$. Here $\mu$ is the energy scale at which these distributions are fitted. Both types of objects are related by
\begin{equation}
H^{i}(x,\xi,t; \mu^2) =
\int_{-1}^{1}d\beta
\int_{-1+|\beta|}^{1-|\beta|}d\alpha \,
\delta(\beta + \xi\alpha - x)
F^{i}(\beta, \alpha, t; \mu^2) \,.
\end{equation}
In the current analysis, we do not include any $D$-term~\cite{Polyakov:1999gs}, as the available data lack sensitivity to the real parts of the amplitudes. In the following equations, for brevity, we will suppress the scale dependence.

The double distributions are expressed as products of generalized $t$-dependent PDFs, $f^A_{i}(\beta, t)$, and a profile function, $h_{i}(\beta, \alpha)$, which generates the dependence on the skewness variable $\xi$:
\begin{equation}
F^{i}(\beta, \alpha, t) = f^A_{i}(\beta, t)\,h_{i}(\beta, \alpha) \,.
\label{eq:dd_master}
\end{equation}
We stress that $H^{i}(x,\xi,t)$ denotes the total contribution of $i$ parton to helium-4. In particular, for $x>0$ the so-called forward limits of our models for quarks, $q = \{u_{\mathrm{val}}, u_{\mathrm{sea}}, d_{\mathrm{val}}, d_{\mathrm{sea}}, s\}$, are 
\begin{equation}
H^{q}(x, 0, 0) = f^A_q(x) \,,
\label{eq:Hx00q}
\end{equation}
which represents the parton distribution functions describing the probability of finding a quark carrying a given fraction of the helium-4 momentum (rather than a given fraction of the proton-in-helium-4 momentum). The corresponding forward limit for gluons is
\begin{equation}
H^{g}(x, 0, 0) = x f^A_g(x) \,.
\label{eq:Hx00g}
\end{equation}
We also note that this work employs the full evolution equations for GPDs, rather than the so-called forward evolution governed by PDFs that is often customary in phenomenological analyses. For this purpose, we utilize the \texttt{APFEL++} package~\cite{Bertone:2022frx, Bertone:2017gds}, which implements the evolution equations for GPDs at LO.

The generalized PDF is
\begin{equation}
f^A_{i}(\beta, t) = f^A_{i}(\beta)\, \frac{k_{i}(|\beta|, t)}{k_{i}(|\beta|, 0)} \,,
\end{equation}
where $f_{i}(\beta)$ accounts for both partons ($\beta > 0$) and antipartons ($\beta < 0$), and we have
\begin{itemize}
\item for $A|\beta| \le 1$
\begin{align}
f^A_{i}(\beta) = A^2
\begin{cases}
\Theta(\beta)\,f^{p/A}_{i}(A|\beta|) \,,
& \mathrm{for}~i = u_{\mathrm{val}}, d_{\mathrm{val}} \,, \\
\mathrm{sgn}(\beta)\,f^{p/A}_{i}(A|\beta|) \,, & \mathrm{for}~i = u_{\mathrm{sea}}, d_{\mathrm{sea}}, s \,, \\
|\beta|\,f^{p/A}_{i}(A|\beta|) \,, & \mathrm{for}~i = g \,,
\end{cases}
\label{eq:forwardFromPheno}
\end{align}
\item otherwise 
\begin{align}
f^A_{i}(\beta) = 0 \,.
\label{eq:forwardFromPhenoZero}
\end{align}
\end{itemize}
Here, $\Theta(\cdot)$ and $\mathrm{sgn}(\cdot)$ are the step and signum functions, respectively, 
while $f^{p/A}_{i}(A|\beta|)$ are PDFs of valence and sea quarks, and of gluons in the nucleon in helium-4. In this analysis we have chosen to use nNNPDF30 parameterizations~\cite{AbdulKhalek:2022fyi}. Since these parameterizations are defined with respect to proton momentum, we rescale $|\beta|$ by $A$. 
Furthermore, since they give parton densities per nucleon, in order to ensure the correct number of valence quarks per helium-4 \mbox{nucleus, i.e.}
\begin{equation}
\int_{-1}^{1}d\beta\,f^A_{u_{\mathrm{val}}}(\beta) = 2 Z + (A-Z) = 6\,,
\end{equation}
and 
\begin{equation}
\int_{-1}^{1}d\beta\,f^A_{d_{\mathrm{val}}}(\beta) = Z + 2 (A-Z) = 6 \,,
\end{equation}
in Eq.~\eqref{eq:forwardFromPheno} we include the prefactor $A$ (another power of $A$ due to the rescaling of the arguments). To avoid probing the PDFs in unknown domains, we explicitly set $f^A_{i}(\beta)=0$ for $\beta>1/A$.

\begin{figure}[!ht] 
  \centering
  \includegraphics[width=0.8\textwidth]{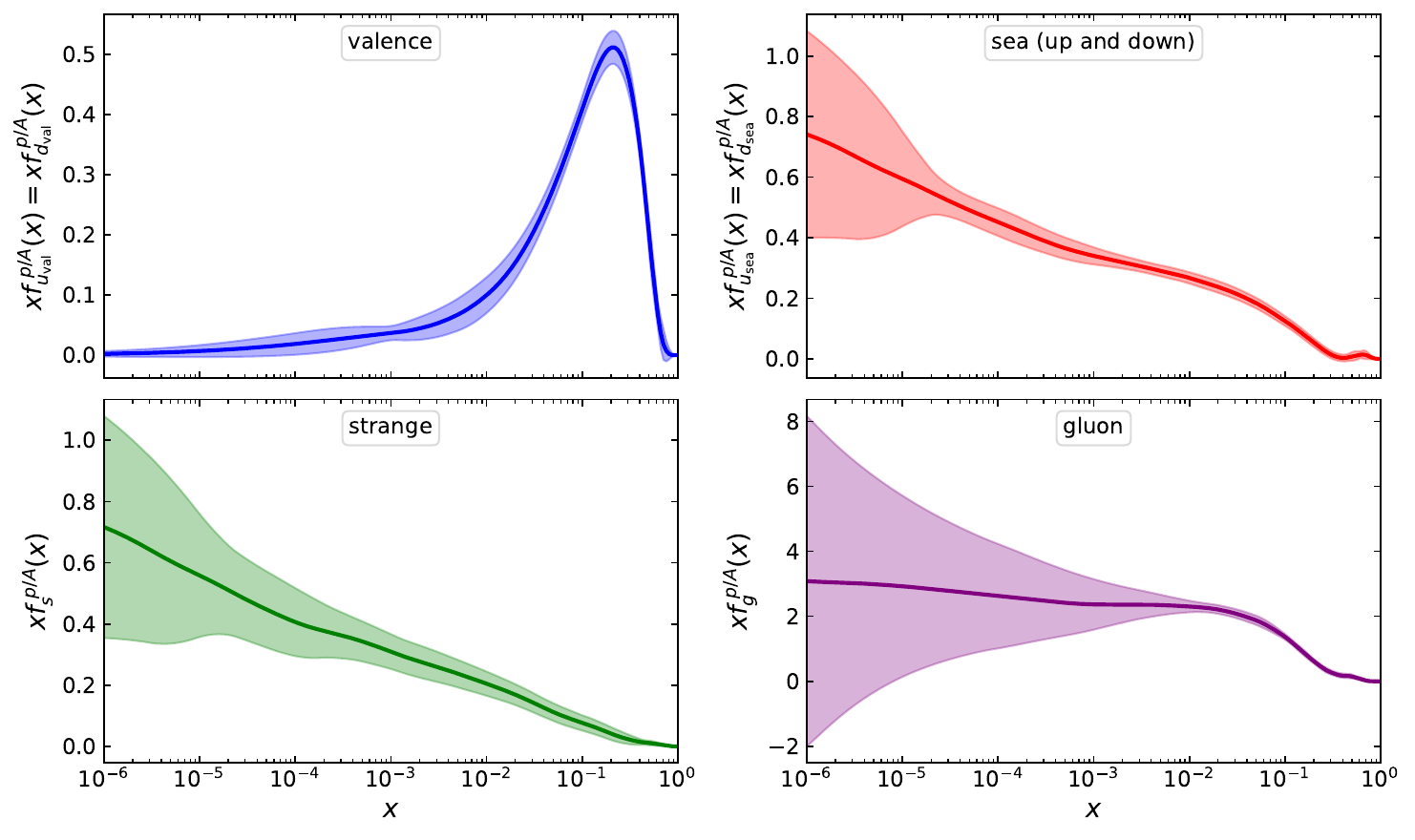}
  \caption{Parameterizations of PDFs for helium-4 obtained by NNPDF group~\cite{AbdulKhalek:2022fyi} for $\mu^2 = 2\,\mathrm{GeV}^2$.}
  \label{fig:pdf}
\end{figure}

\begin{figure}[!ht] 
  \centering
  \includegraphics[width=0.4\textwidth]{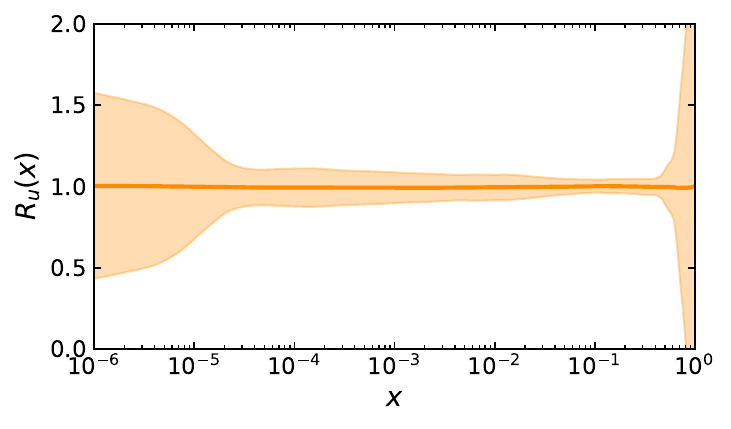}
  \caption{The nuclear modification ratio $R_q(x)$, see Eq.~\eqref{eq:pdfR}, for up quarks and $\mu^2 = 2\,\mathrm{GeV}^2$.}
  \label{fig:R}
\end{figure}

Figure~\ref{fig:pdf} shows the nNNPDF30 parametrizations of the helium-4 PDFs, while Fig.~\ref{fig:R} shows the nuclear modification ratio, defined as
\begin{equation}
R_{i}(x) = \frac{A\,f_{i}^{p/A}(x)}{Z\,f_{i}^{p}(x) + (A-Z)\,f_{i}^{n}(x)} \,.
\label{eq:pdfR}
\end{equation}
Here, $f_{i}^{p}(x)$ and $f_{i}^{n}(x)$ denote the PDFs of the proton and neutron, respectively, coming from the same set of phenomenological solutions, and no distinction is made between valence and sea contributions. 

In this analysis, we refrain from using fully factorized dependencies in $x$ and $t$. Instead, for valence quarks, $i = \{u_{\mathrm{val}}, d_{\mathrm{val}}\}$, we have
\begin{align}
\displaystyle{
k_{i}(|\beta|, t) = 
\left(\frac{1}{1-p_0(1-|\beta|)^2t}\right)^{p_1}
\prod_{j=1}^{n}\left(\frac{|p_{2,j} + t|}{|p_{2,j} + t| - p_{3,j}(1-|\beta|)^2t}\right)^{p_{4,j}}
}\,.
\label{eq:tDep}
\end{align}
This Ansatz consists of two distinct components. The first, a dipole-like Ansatz, captures the main trend in $t$ for a given $x$. The second, represented by the product, accounts for $n$ diffractive minima occurring at $-t=\{p_{2, 1}, p_{2, 2}, \ldots\}$.
Our Ansatz ensures that.
\begin{align}
\lim_{|\beta| \to 1} k_{i}(|\beta|, t) = \mathrm{const.} \,,
\end{align}
which reflects a theory-driven constraint arising from hadron tomography. Specifically, in the limit $x = 1$, where the entire momentum of the hadron is carried by a single parton, the position of this parton coincides with the origin of the coordinate system in which the hadronic tomography is defined.

The free parameters of Eq.~\eqref{eq:tDep} are fixed by a fit to the helium-4 elastic form factor data, as detailed in the Sec.\ref{sec::Results}, using the relation
\begin{equation}
F(t) = \frac{1}{Z} \int_{-1}^{1}dx \left(e_{u}f_{u_{\mathrm{val}}}^A(x, t) + e_{d}f_{d_{\mathrm{val}}}^A(x, t)\right) \,,
\label{eq:eff}
\end{equation}

For sea quarks and gluons,  
$i = \{u_{\mathrm{sea}}, d_{\mathrm{sea}}, s, g\}$, we use a different Ansatz, namely 
\begin{align}
\displaystyle{
k_{i}(|\beta|, t) = 
\exp(p(1-|\beta|^2)t)
}\,.
\label{eq:tDepSea}
\end{align}
The slope $p$ will be fitted to CLAS data~\cite{CLAS:2017udk}, as described in the Sect.~\ref{sec::Results}. 

For the profile function in Eq.~\eqref{eq:dd_master}, we use the classic Radyushkin's Ansatz,
\begin{equation}
h_{i}(\beta, \alpha) = 
\frac{\Gamma(2b_{i}+2)}{2^{2b_{i}+1}\Gamma^2(b_{i}+1)}
\frac{\left( (1-|\beta|)^2 - \alpha^2\right)^{b_{i}}}{(1-|\beta|)^{2b_{i}+1}}\,,
\label{eq:dd_profile}
\end{equation}
whose normalization
\begin{equation}
\int_{-1+|\beta|}^{1-|\beta|}d\alpha\,h_{i}(\beta, \alpha) = 1\,,
\end{equation}
ensures the desired reduction at $\xi = 0$, already expressed in Eqs.~\eqref{eq:Hx00q} and~\eqref{eq:Hx00g}.

The coefficients controlling the strength of the skewness effect are $b_{i} = 1$ for valence quarks and $b_{i} = 2$ for sea quarks and gluons, which are typical values adopted in GPD models for the proton, for instance in Ref.~\cite{Goloskokov:2006hr}. 
 
For the purpose of our analysis, we also need a model for $H_{T}^{g}$, which contributes at NLO to the $\amp^{+-}$ amplitude. Since little is known about this GPD, we bound it by $H^g(x,\xi,0)$ as $t \to 0$, while for $t \ll 0$ we use theory-driven constraints, namely the positivity inequality~\cite{Pire:1998nw} derived in Ref.~\cite{Kirch:2005in} for $x > \xi$:
\begin{equation}
    |H_T^g (x,\xi,t)|<
    \frac{1}{\xi}
    \sqrt{\frac{x^2-\xi^2}{1-\xi^2}}
    \sqrt{\frac{t}{t-t_0}} 
    \sqrt{\frac{t_0}{t-t_0}} \sqrt{g\left(\frac{x+\xi}{1+\xi}\right) g\left(\frac{x-\xi}{1-\xi}\right)} \,,
    \label{eq:HtPos}
\end{equation}
where
\begin{equation}
    t_0 = -4 M^2 \frac{\xi^2}{1-\xi^2}\,.
\end{equation}
We note that
\begin{equation}
    \sqrt{\frac{t}{t-t_0}} 
    \sqrt{\frac{t_0}{t-t_0}} = 
    2\, M\, \xi\sqrt{-1/t} + O(\xi^3) \,,
\end{equation}
so for $\xi \to 0$
\begin{equation}
    |H_T^g (x,\xi \to 0,t)|<
    x\,g(x)
    \sqrt{\frac{1}{-t/(4M^2)}}\,.
    \label{eq:HtPosxi0}
\end{equation}
This suggests the following Ansatz, which we adopt in this analysis:
\begin{equation}
    H_{T}^{g}(x, \xi, t) = H^{g}(x, \xi, 0)\sqrt{\frac{1}{1 - t/(4M^2)}}\,.
\end{equation}
The main feature of this solution is its conservativeness: at $t=0$, we find $H_{T}^{g}(x, \xi, 0) = H^{g}(x, \xi, 0)$, whereas for $|t| > 0$, $H_{T}^{g}(x, \xi, t)$ exhibits a much milder dependence on $t$ than $H^{g}(x, \xi, t)$.

It remains instructive to verify whether our model for $H_{T}^{g}$, now based on the 
double distribution representation, fulfills the positivity constraint 
for $\xi > 0$. 

\begin{figure}[!ht] 
  \centering
  \includegraphics[width=0.6\textwidth]{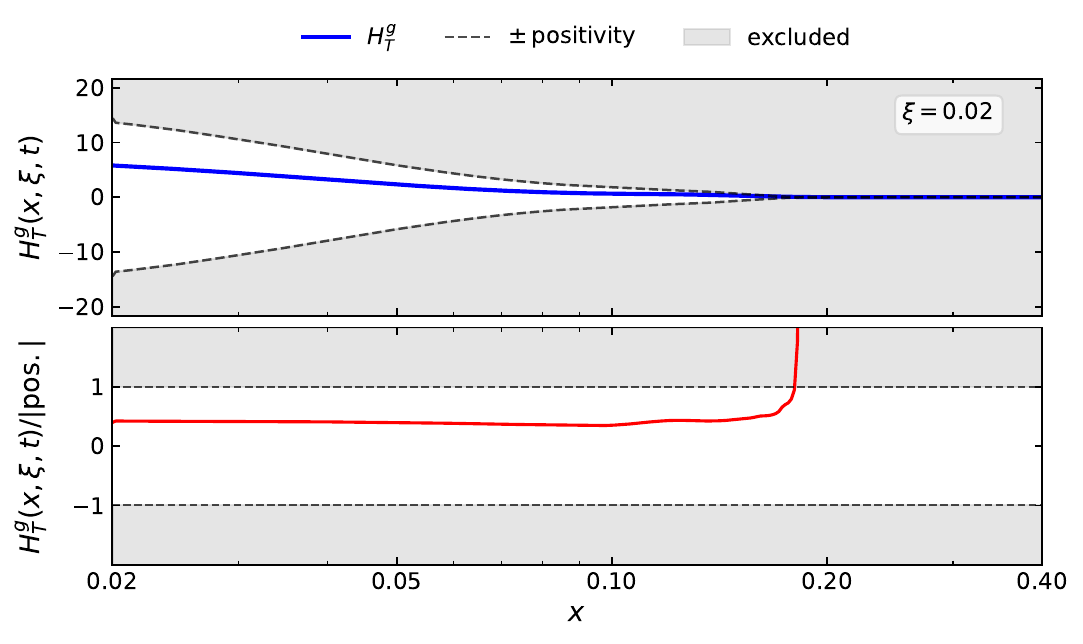}
  \caption{Our model for the GPD $H_{T}^{g}$ compared against the positivity bound, see Eq.~\eqref{eq:HtPos}, for $\xi=0.02$ and $t=-10\,\mathrm{GeV}^2$.}
  \label{fig:pos}
\end{figure}

The results are illustrated in 
Fig.~\ref{fig:pos}, which displays our model against the positivity 
bound. We observe that the constraint is only violated in the large-$x$ 
tail of the distribution. This is expected, as in our analysis the gluon PDF for helium-4 vanishes 
identically for $x > 1/A$, see Eq.~\eqref{eq:forwardFromPhenoZero}. Consequently, the right-hand side of Eq.~\eqref{eq:HtPos} always vanishes 
for $x > (1 + \xi(1-A))/A$, whereas the left-hand side retains a 
non-zero dependence due to the inherent properties of the Radon transform. 
We note that since our phenomenological analysis focuses on the 
small-$\xi$ region where these effects are suppressed, this localized 
violation at large $x$ does not undermine the overall validity of our results.

As the positivity constraint does not fix the sign of the GPD $H_{T}^{g}$, it would seem reasonable to vary it in the phenomenological analysis. We, however, refrain from doing so, as we find the contribution from $H_{T}^{g}$, despite taking a conservative estimate, to be negligible (see the next section). 


\section{Results}
\label{sec::Results}

In this section, we collect the main results of our analysis of helium-4 GPDs. We first determine the model parameters from elastic form factor and beam-spin asymmetry data, and then discuss the resulting tomographic picture. We also examine the impact of NLO and kinematic higher-twist effects on the description of present data and on predictions for future JLab12 observables.

Before turning to the results in detail, we briefly comment on how the extraction framework used in this work is related to the kinematics employed for tomography. In the present work, the GPDs are extracted using the approach of Refs.~\cite{Braun:2022qly,Martinez-Fernandez:2025gub}, which is particularly well suited for the systematic organization of kinematic higher-twist corrections. When discussing tomography, however, we employ the usual symmetric frame~\cite{Diehl:2003ny}, in which the $\xi=0$ limit corresponds to a purely transverse momentum transfer, $t=-\boldsymbol{\Delta_T}^2$, thus recovering the conventional impact-parameter representation for densities:
\begin{equation}
f^A_i(x,\boldsymbol{b_T}) =
\int \frac{\mathrm{d}^2\boldsymbol{\Delta_T}}{(2\pi)^2}  
e^{-i\boldsymbol{b_T}\cdot \boldsymbol{\Delta_T}}
H^i(x,0,-\boldsymbol{\Delta_T}^2) \,.
\end{equation}
The impact parameter, $\boldsymbol{b_T}=(b_x, b_y)$, is defined in a coordinate system whose origin is set at the center of momentum of all proton constituents~\cite{Burkardt:2002hr}.

\begin{figure}[!htbp] 
  \centering
  \includegraphics[width=0.4\textwidth]{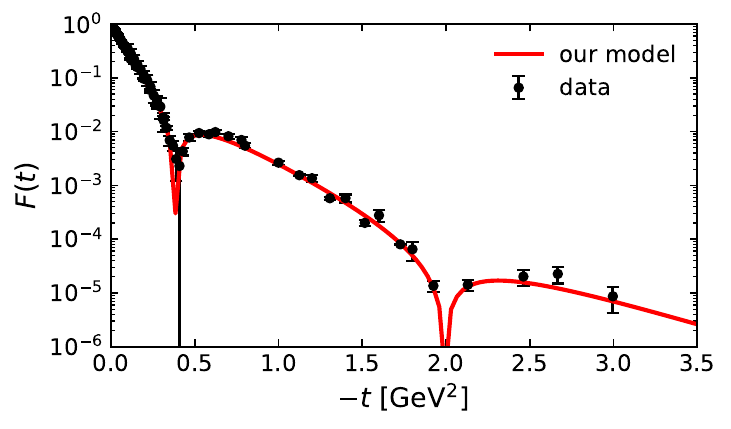}
  \includegraphics[width=0.4\textwidth]{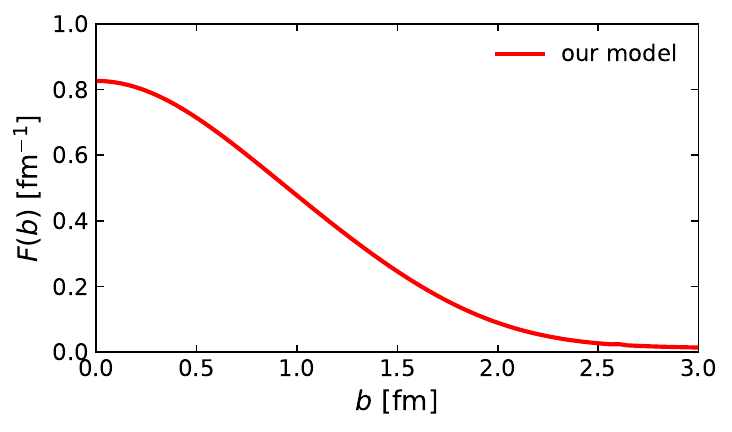}
  \caption{(left) Fit to helium-4 elastic form factor data. (right) Fourier transform of fitted helium-4 elastic form factor, where $\int_0^{\infty}db\,F(b) = 1$.}
  \label{fig:eff_fit}
\end{figure}

As a first step of our fit procedure, we determine the free parameters of Eq.~\eqref{eq:tDep} from the helium-4 elastic form factor data. We model the first two diffractive minima ($n=2$). The fit, based on the data from Refs.~\cite{Repellin:1965vba,Frosch:1967pz,Arnold:1978qs,Ottermann:1985km,JeffersonLabHallA:2013cus}, is shown in Fig.~\ref{fig:eff_fit}, together with the corresponding spatial charge distribution in helium-4, obtained by Fourier transforming the fitted form factor to impact-parameter space. No constraints are imposed on the fit parameters, and we obtain:
$p_0 =	(0.088 \pm 0.016)/\mathrm{GeV}^2$,
$p_1=	39.5 \pm 6.4$,
$p_{2,1}=	(0.3833 \pm 0.0024)\,\mathrm{GeV}^2$,
$p_{3,1}=	3.49 \pm 0.16$,
$p_{4,1}=	0.985 \pm 0.027$,
$p_{2,2}=	(1.993 \pm 0.018)\,\mathrm{GeV}^2$,
$p_{3,2}=	2.38 \pm 0.86$ and
$p_{4,2}=	0.988 \pm 0.099$.
Due to the isospin symmetry, we use the same parameters for valence up and down quarks. 

The last remaining parameter of the model, namely the slope $p$, is fitted to the CLAS data~\cite{CLAS:2017udk}, for the beam-spin asymmetry, defined as
\begin{equation}
    A_{LU}(\phi) =
    \frac{
    \sigma^{++}_- +
    \sigma^{+-}_- -
    \sigma^{-+}_- -
    \sigma^{--}_-
    }
    {
    \sigma^{++}_- +
    \sigma^{+-}_- +
    \sigma^{-+}_- +
    \sigma^{--}_-
    }
\end{equation}
where the differential cross section,  $\sigma^{s\lambda}_\chi$, for a beam of charge sign $\chi$ and helicity $s$, producing a real photon with polarization $\lambda$ was defined in Eq.~\eqref{cross_section}. 

We perform three distinct fits, calculating asymmetries using LO/LT, NLO/LT and NLO/HT coefficient functions. Each calculation relies on specific types of GPDs, yields distinct amplitudes, and leads to varying levels of agreement with the experimental data, which we summarize in Table~\ref{tab:summary}. We note that in the fit utilizing LO/LT coefficient functions, the $\chi^2$ function does not reach a minimum. Instead, it saturates for large values of $p$. Therefore, we are only able to provide a lower limit for this parameter. This indicates that when using LO/LT coefficient functions, the CLAS data prefers to have the sea contribution totally suppressed. 
\begin{table}[!htbp] 
\caption{Comparison of our $A_{LU}$ asymmetry estimates with CLAS data~\cite{CLAS:2017udk}. The three rows represent fits using different coefficient functions, corresponding to varying levels of precision in the $\alpha_s$ and kinematic twist expansions. These fits utilize the specific GPD types and amplitudes listed in the table. The agreement between constrained models and the experimental data is quantified by the global $\chi^2/n$ value, where $n$ is the number of points measured.}
\label{tab:summary}
\begin{ruledtabular}
\begin{tabular}{cccccc}
\multicolumn{2}{c}{\hspace{-20mm}Precision in} & \multirow{2}{*}{GPDs involved} & \multirow{2}{*}{Non-vanishing amplitudes} & Fitted value of $p$ & \multirow{2}{*}{Global $\chi^2/n$} \\$\alpha_s$ expansion & kinematic twist expansion & & &  $[\mathrm{GeV}^{-2}]$ & \\ 
\colrule
LO & LT & $H^q$ & $\amp^{++}$ & $ >70$ & 1.31 \\ 
NLO & LT & $H^q$, $H^g$, $H_{T}^g$ & $\amp^{++}$, $\amp^{+-}$ & $23.4_{-3.9}^{+7.4}$ & 1.28 \\
NLO & HT & $H^q$, $H^g$, $H_{T}^g$ & $\amp^{++}$, $\amp^{+-}$, $\amp^{0+}$ & $22.0^{+5.5}_{-2.1}$ & 1.00 \\
\end{tabular}
\end{ruledtabular}
\end{table}

For our nominal fit utilizing the NLO/HT coefficient functions we obtain $p=22.0^{+5.5}_{-2.1}~\mathrm{GeV}^{-2}$. Such a value implies a steeper dependence on $t$ than in the valence sector and, consequently, a broader spatial distribution of sea quarks and gluons, as expected. More robust constraints on the tomography of sea quarks and gluons should become possible with upcoming JLab data and, in the longer term, with data from the EIC. 

\begin{figure}[!ht] 
  \centering
  \includegraphics[width=0.9\textwidth]{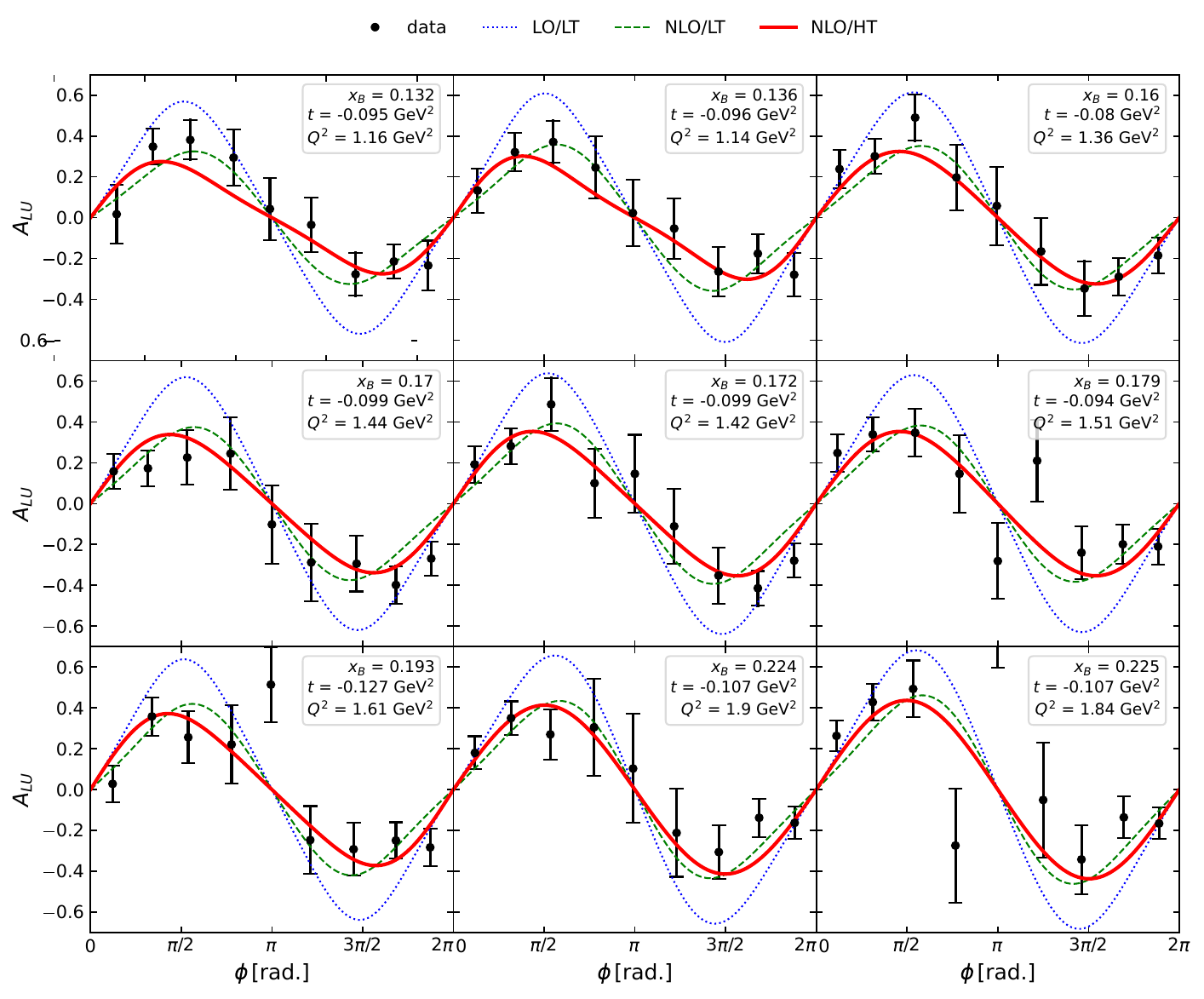}
  \caption{Experimental data for the $A_{LU}$ asymmetry~\cite{CLAS:2017udk} compared with our model. See the legend at the top of the figure for more information.}
  \label{fig:clas_alu}
\end{figure}

Figure~\ref{fig:clas_alu} compares the $A_{LU}$ asymmetries measured by CLAS~\cite{CLAS:2017udk} with our predictions. For each $(x_{B}, t, Q^2)$ kinematic bin, three curves are shown, obtained from the same underlying GPD parametrization but using coefficient functions evaluated at LO/LT, NLO/LT, and NLO/HT accuracy.

\begin{figure}[!ht] 
  \includegraphics[width=0.8\textwidth]{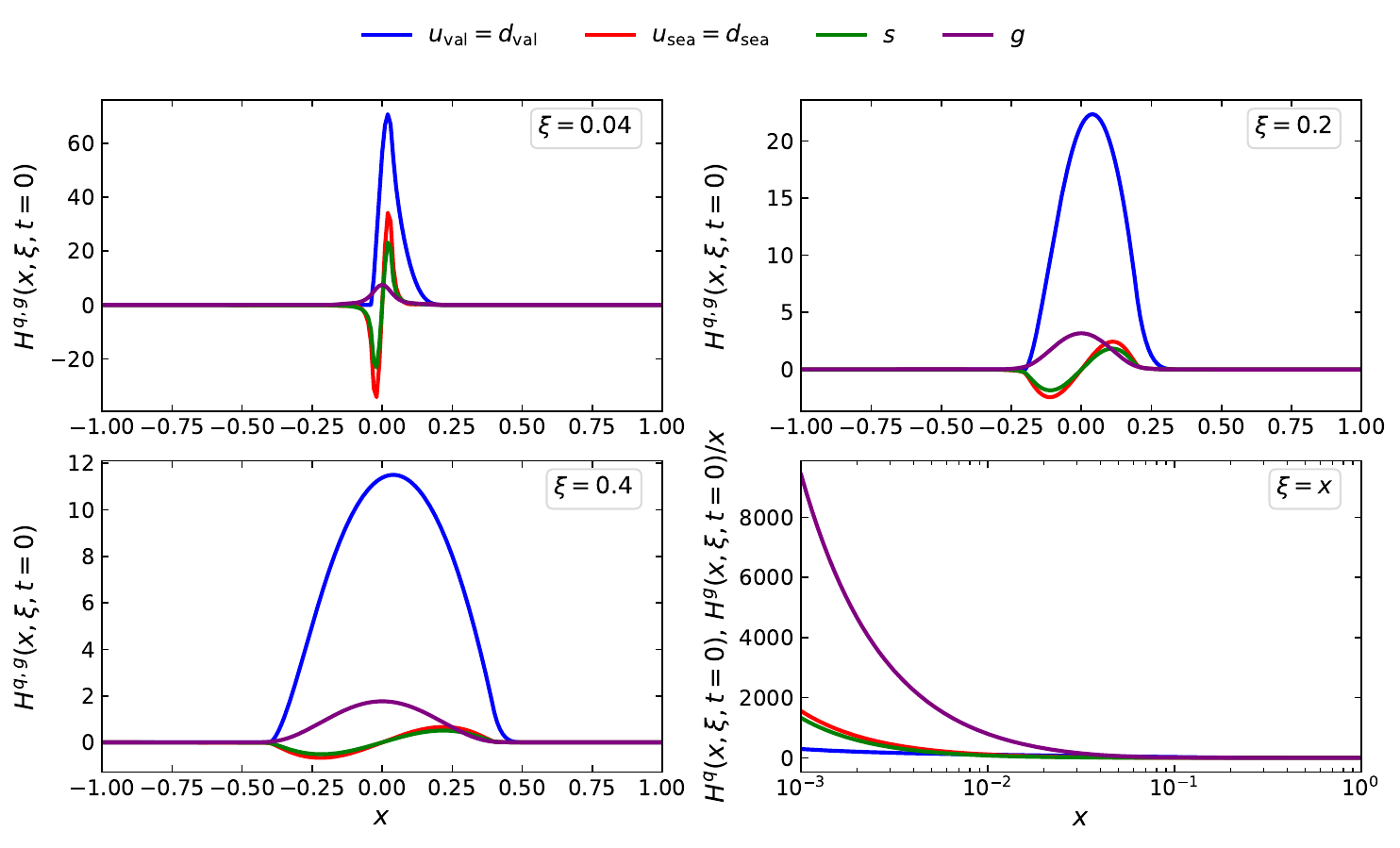}
  \caption{Dependence of our GPD models for helium-4 on the variable $x$ for $\xi = \{0.04,0.2,0.4,x\}$ at $|t| = 0.2\,\mathrm{GeV}^2$ and $\mu^2 = 2\,\mathrm{GeV}^2$. In the plot for $\xi=x$ the gluon GPD is divided by $x$.}
  \label{fig:xDep}
\end{figure}

\begin{figure}[!ht] 
  \centering
  \includegraphics[width=0.9\textwidth]{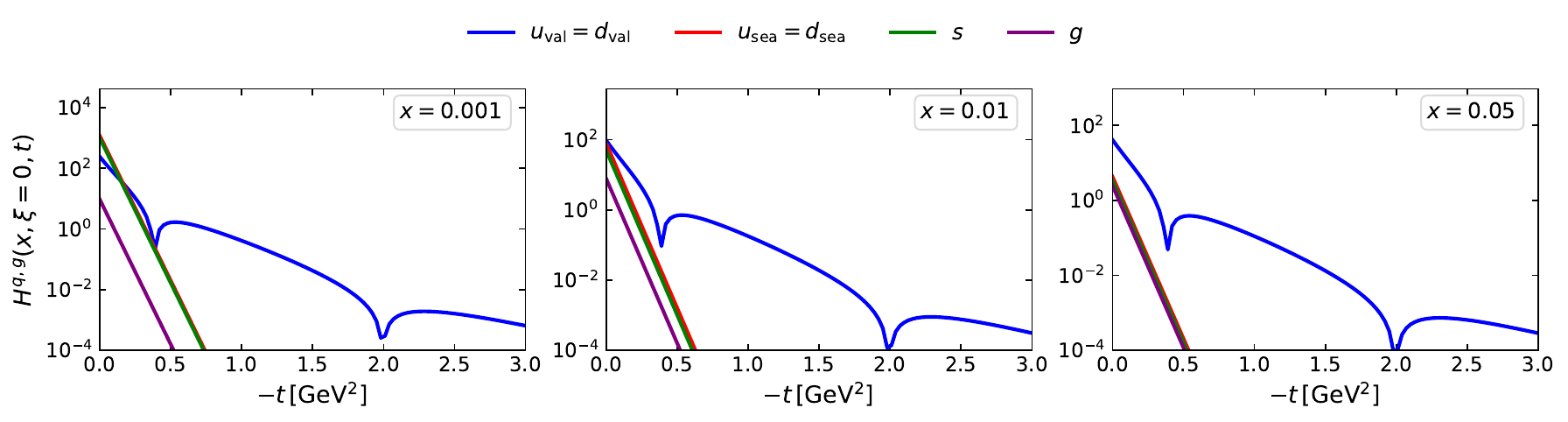}
  \includegraphics[width=0.9\textwidth]{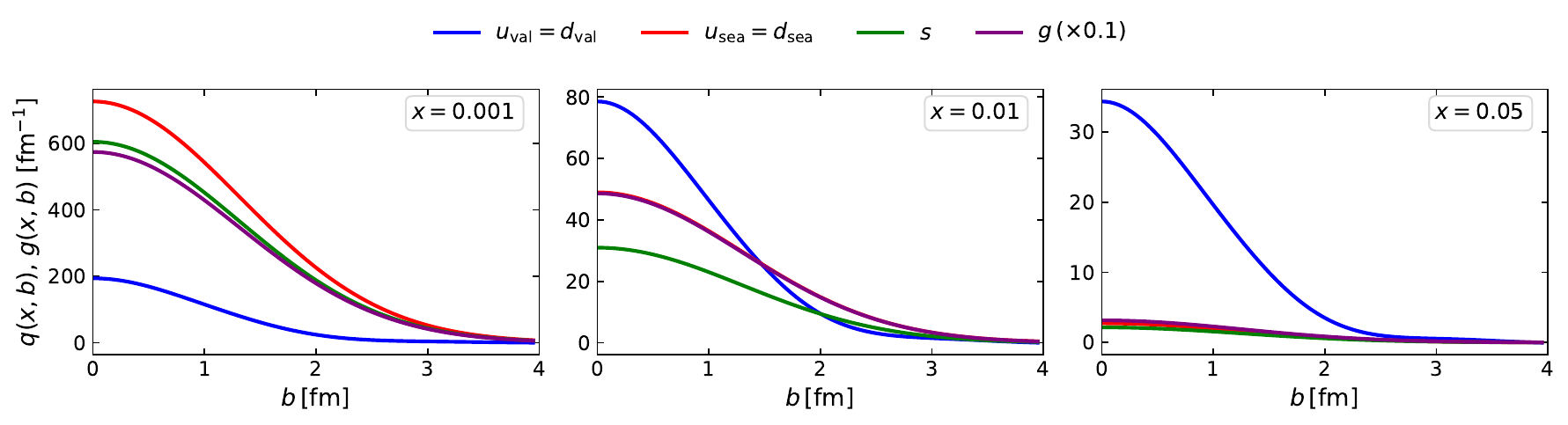}
  \caption{Dependence of our GPD models for helium-4 on the variable $t$ for $\xi=0$, $\mu^2 = 2\,\mathrm{GeV}^2$ and $x = \{0.001,0.01,0.05\}$, and corresponding tomographic pictures, where $\int_0^{\infty}db\,q(x, b) = q(x)$ and similar for gluons. In the lower plots, the gluon curves are scaled by a factor of $0.1$.}
  \label{fig:nt}
\end{figure}

To illustrate the main properties of the fitted model, we show its dependence on the relevant kinematic variables in the following figures. The dependence of our model on the variable $x$ is shown in Fig.~\ref{fig:xDep} for several values of $\xi$. For small skewness, the support is similar to that of an ordinary PDF, with the distributions vanishing for $x \gtrsim 0.25$, while for larger $\xi$ it becomes broader, reflecting the support properties of the underlying double-distribution representation. In Fig.~\ref{fig:nt}, we illustrate the dependence on the variable $t$ at fixed $x$ and $\xi=0$, including the corresponding hadron tomography images. The figure clearly demonstrates the steep fitted slope of the sea-quark and gluon contributions.

\begin{figure}[!ht] 
  \centering
  \includegraphics[width=0.9\textwidth]{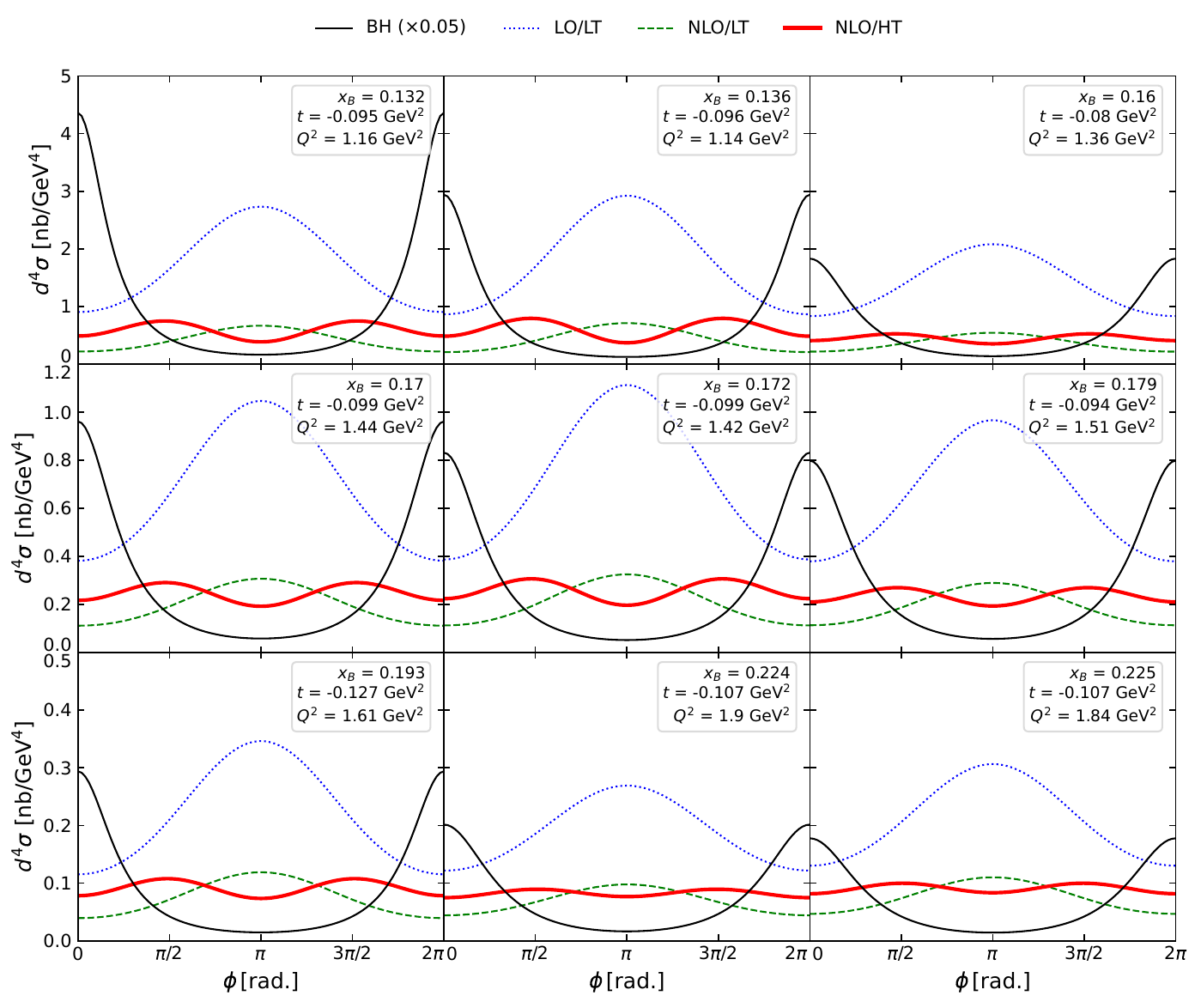}
  \caption{Differential cross-section, $d^4\sigma/(dx_{B}dtdQ^2d\phi)$, for the BH and DVCS sub-processes. For better visibility, the BH curves have been scaled down by a factor of $20$.}
  \label{fig:clas_cs}
\end{figure}

Figure~\ref{fig:clas_cs} shows the unpolarized cross sections corresponding to the kinematic bins used in the CLAS measurement. The figure also includes the contribution from the BH process, which has been scaled down by a factor of $20$ for visual clarity. These plots illustrate the impact of NLO and HT corrections on both the shape and magnitude of the pure DVCS cross section. Consequently, constraining the unpolarized pure DVCS cross section would provide a stringent test of these perturbative frameworks, which is a task that remains challenging due to the BH dominance. 

\begin{figure}[!ht] 
  \centering
  \includegraphics[width=0.7\textwidth]{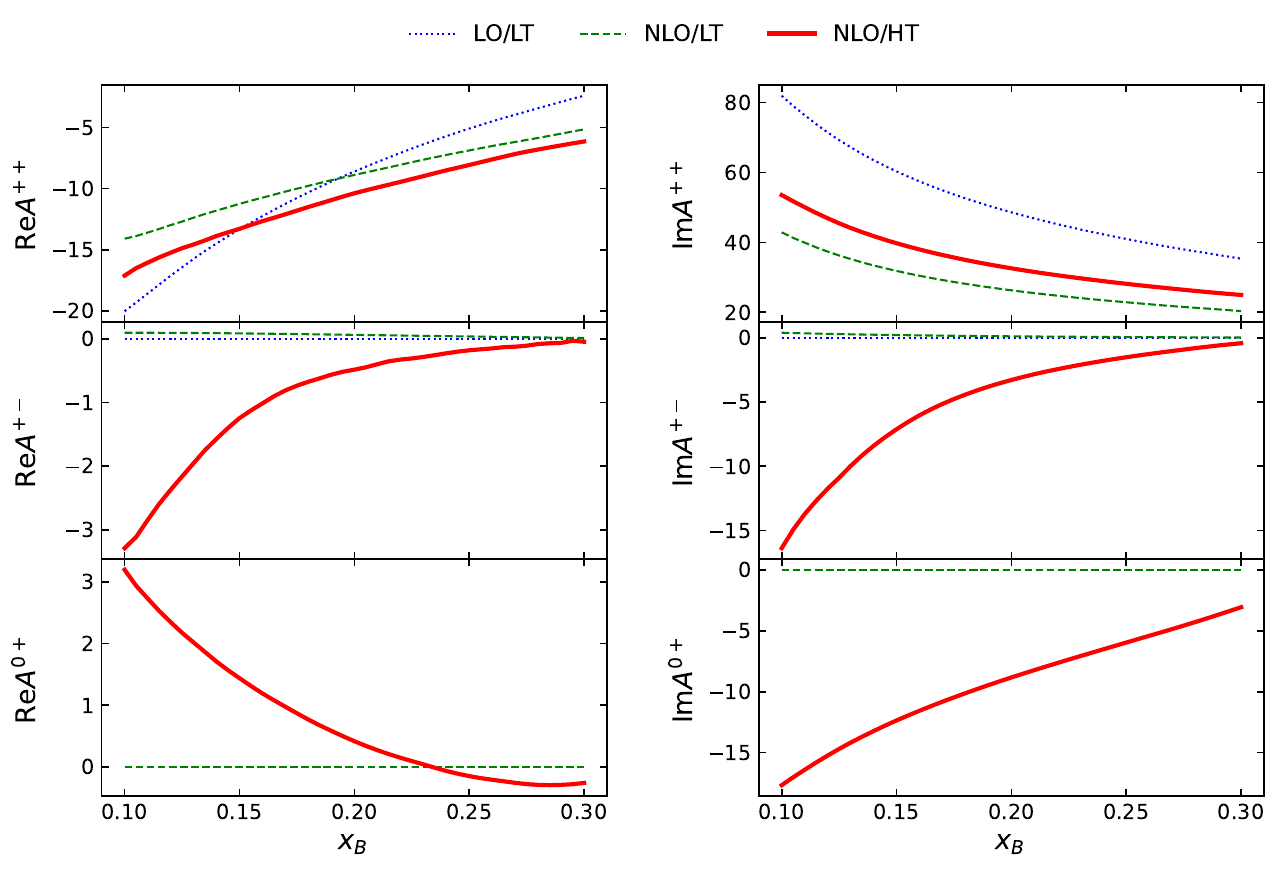}
  \includegraphics[width=0.7\textwidth, trim=0 0 0 1.cm, clip]{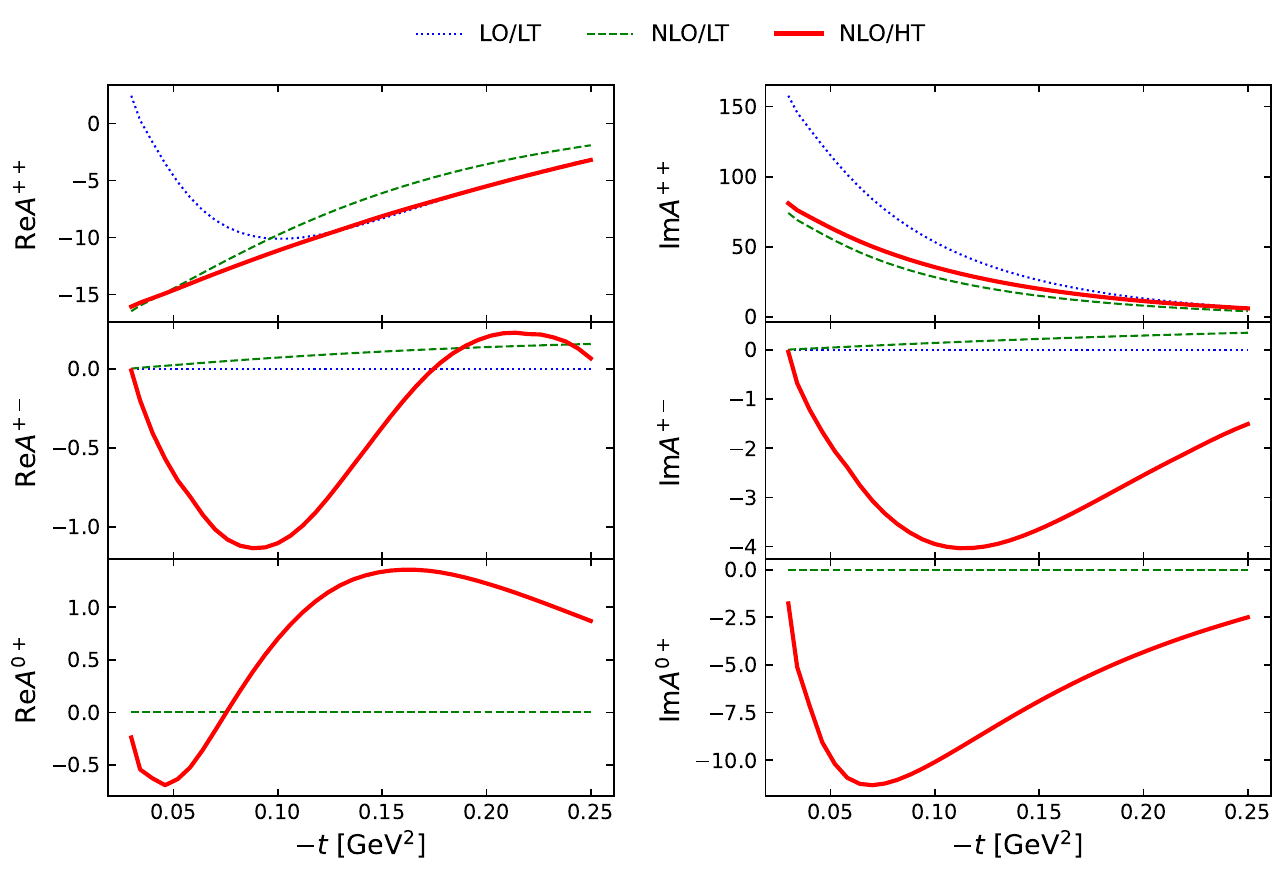}
  \includegraphics[width=0.7\textwidth, trim=0 0 0 1.cm, clip]{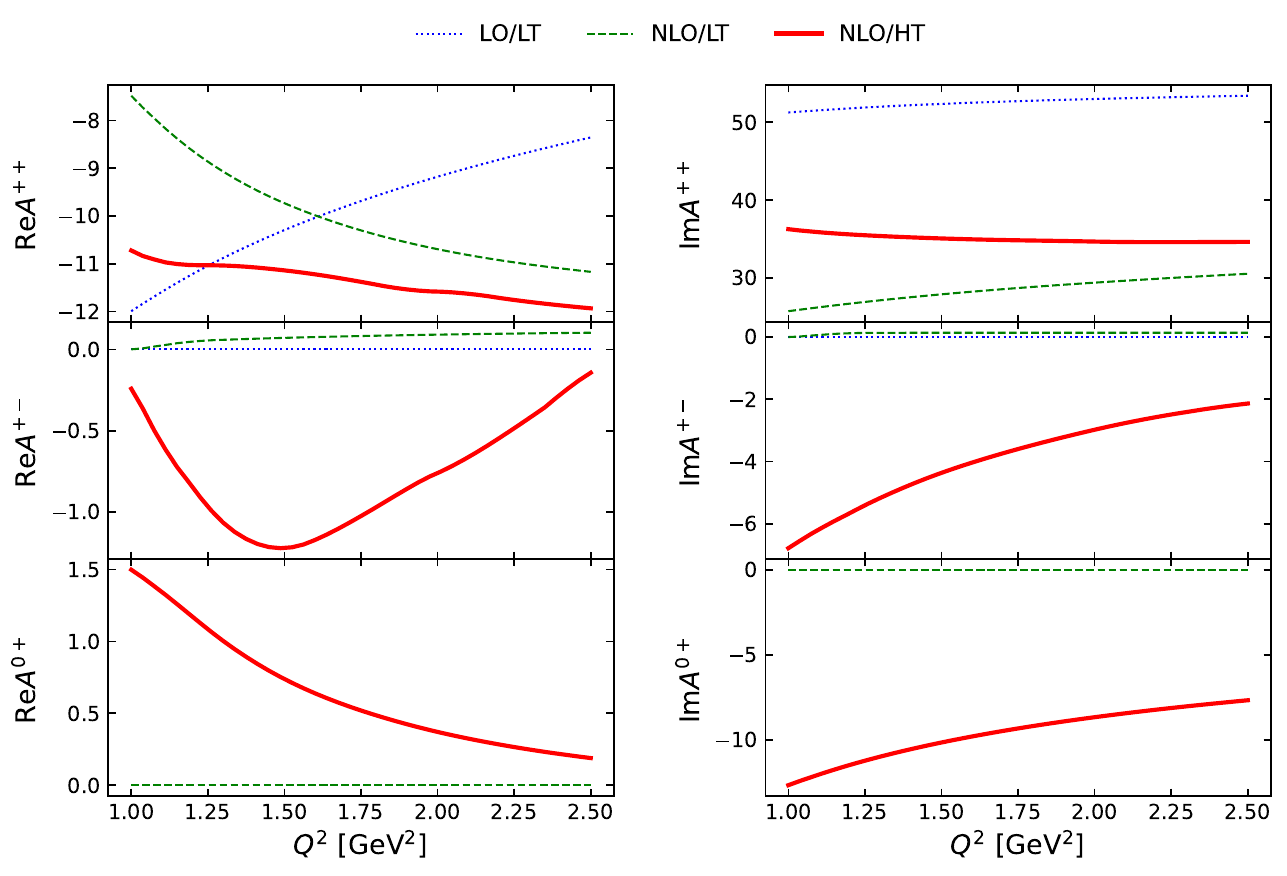}
  \caption{Amplitudes as a function of $x_B$ (top row) for $t = -0.1\,\mathrm{GeV}^2$ and $Q^2 = 1.5\,\mathrm{GeV}^2$; $-t$ (middle row) for $x_B = 0.18$ and $Q^2 = 1.5\,\mathrm{GeV}^2$; and $Q^2$ (bottom row) for $x_B = 0.18$ and $t = -0.1\,\mathrm{GeV}^2$.}
  \label{fig:amps}
\end{figure}

The three helicity amplitudes contributing to the DVCS process, shown as functions of $x_B$, $t$, and $Q^2$ in Fig.~\ref{fig:amps}, clearly exhibit the dominance of the imaginary part of $\amp^{++}$. The figure also confirms the expected pattern of higher-twist (HT) corrections, which become suppressed at large $Q^2$ and small $t$. It furthermore shows that the gluon-transversity contribution to $\amp^{+-}$ remains small compared with the HT contribution.

\begin{figure}[!ht] 
  \centering
  \includegraphics[width=0.5\textwidth]{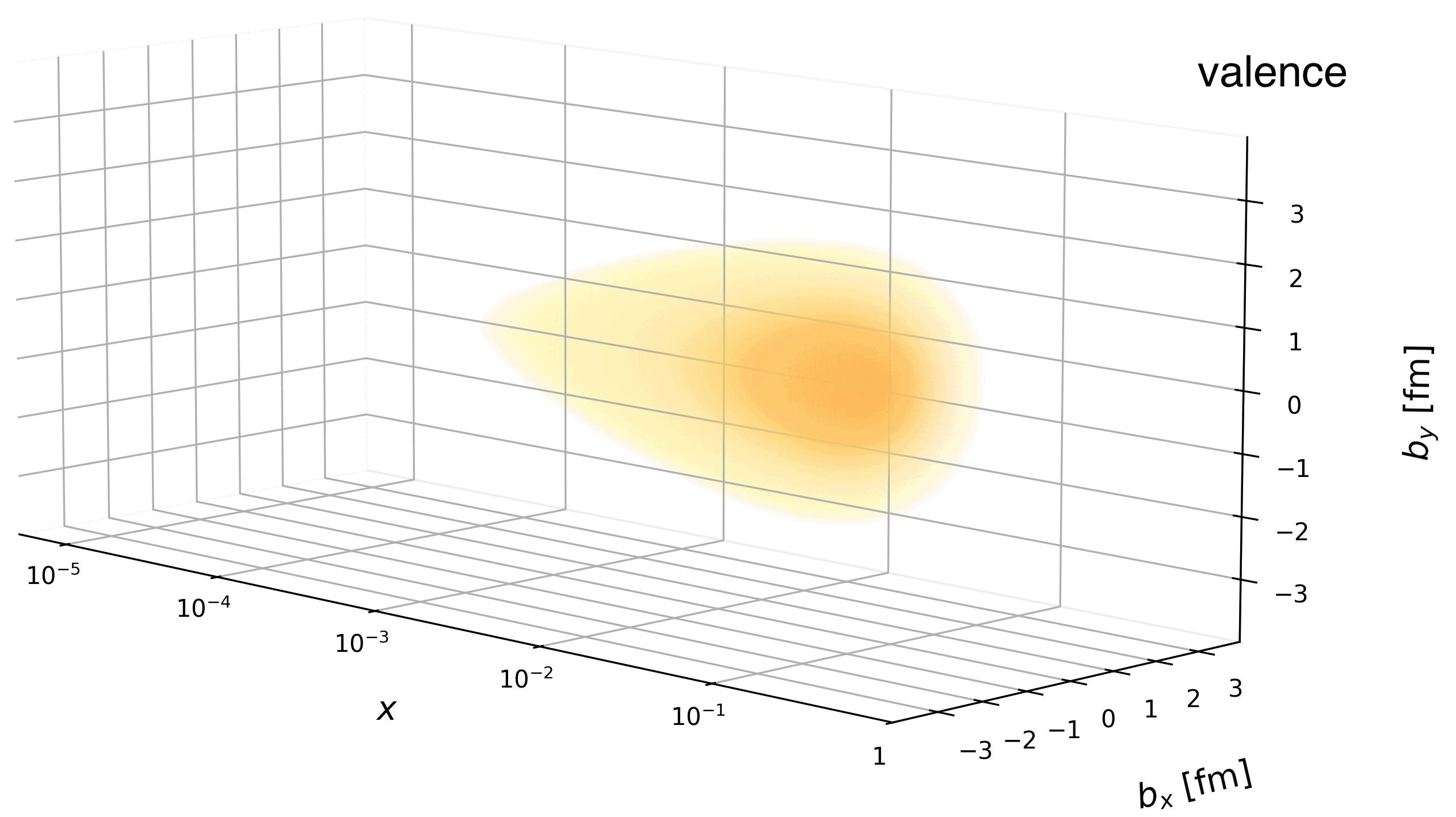}
  \includegraphics[width=0.5\textwidth]{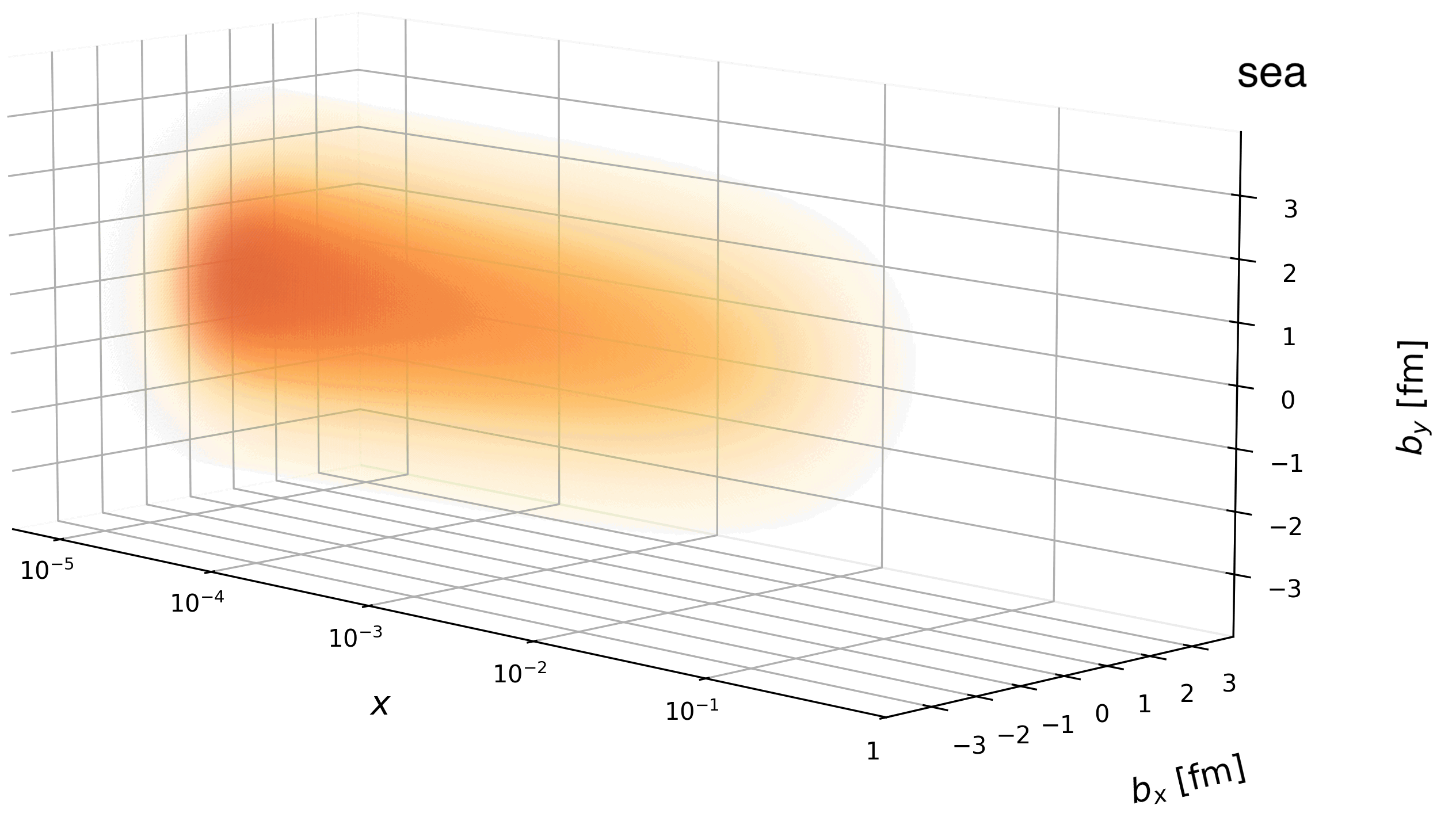}
  \includegraphics[width=0.5\textwidth]{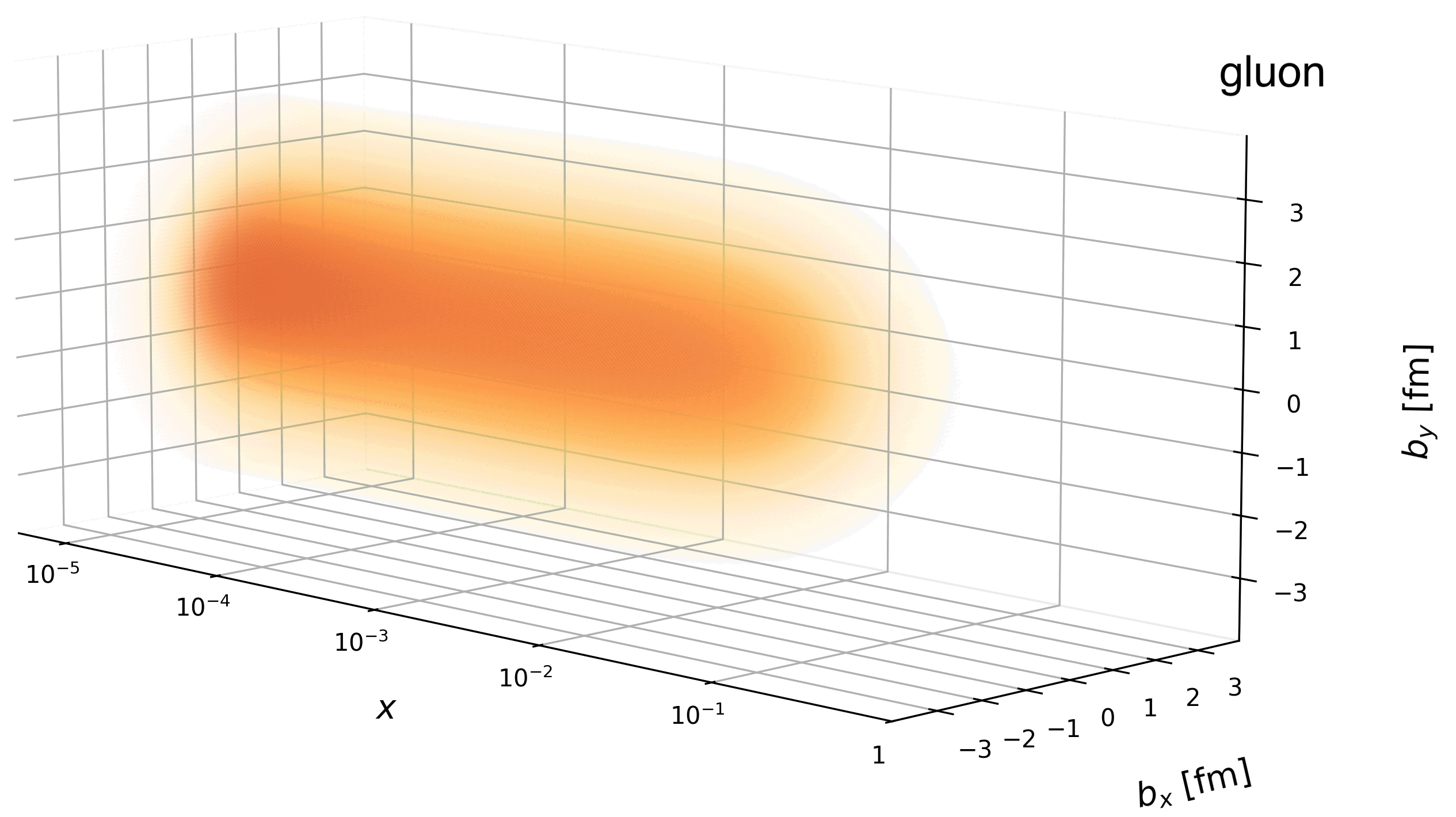}
  \caption{Tomographic pictures of $^4\mathrm{He}$ nuclei, $x\,q(x, b_x, b_y)$ and $x\,g(x, b_x, b_y)$. The valence distribution indicates contributions coming from both up and down quarks, while the sea distribution includes contributions from up, down and strange quarks and antiquarks.}
  \label{fig:nt_3d}
\end{figure}

Figure~\ref{fig:nt_3d} presents 3D tomographic images of the $^4\mathrm{He}$ nucleus, shown separately for valence quarks, sea quarks, and gluons. We see that the valence quarks, on average, carry higher momenta (as expected from PDF analyses) and are embedded within a broader spatial distribution of softer sea quarks and gluons. \footnote{A complementary representation is included in the companion paper~\cite{Martinez-Fernandez:2026}.}

\begin{figure}[!ht] 
  \centering
  \includegraphics[width=0.7\textwidth]{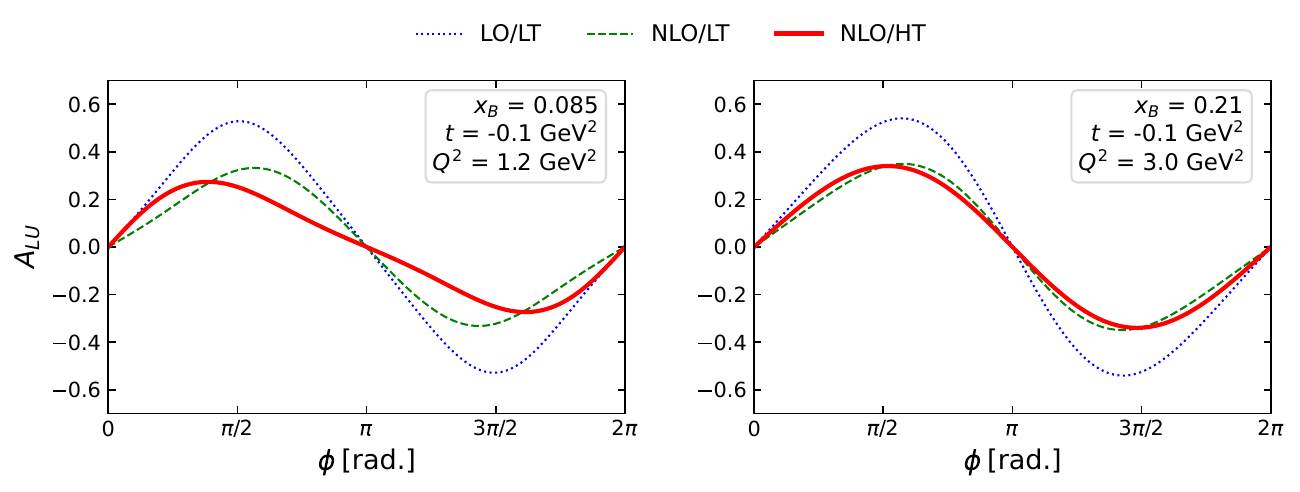}
  \includegraphics[width=0.7\textwidth]{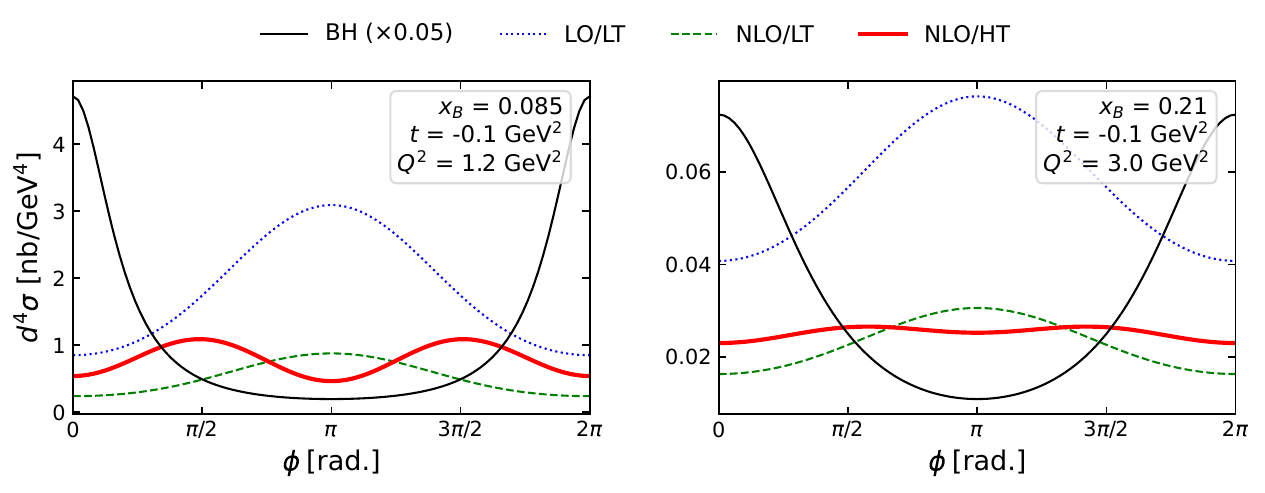}  
  \caption{Predictions for JLab12: differential cross-sections (upper row) and $A_{LU}$ asymmetries (lower row) for $Q^2 = 1.2\,\mathrm{GeV}^2$ (left) and $Q^2 = 3\,\mathrm{GeV}^2$ (right), at $y=0.75$, $t = -0.1\,\mathrm{GeV}^2$, and an electron beam energy of $10.06\,\mathrm{GeV}$.}
  \label{fig:jlab12}
\end{figure}

Finally, Fig.~\ref{fig:jlab12} presents our predictions for the $A_{LU}$ asymmetries and unpolarized cross sections at typical JLab12 kinematic settings. We select two values of $Q^2$, namely $1.2\,\mathrm{GeV}^2$ and $3\,\mathrm{GeV}^2$, while keeping $y$ fixed, demonstrating the suppression of HT effects in the latter case.

\clearpage
\section{Conclusions
}\label{sec::conclusions}
Our phenomenological study of DVCS on the helium-4 nucleus has highlighted the importance of including both higher-order corrections in $\alpha_s$ and kinematical twist-3 and twist-4 contributions to the QCD amplitude. This improved level of accuracy allowed for a successful description of the beam-spin asymmetry data and for constraining the gluon, valence-quark, and sea-quark GPDs of the helium-4 nucleus. It has also demonstrated that a first analysis of the impact-parameter dependence of GPDs is already possible, thereby opening the way to a first quark and gluon tomography of a light nucleus. More generally, our results indicate that such an extended treatment is likely to be fruitful, and perhaps even necessary, also in phenomenological analyses of Compton processes on the proton and neutron. At the same time, the role of genuine twist-3 and twist-4 effects remains an open theoretical issue that must be clarified before fully definitive conclusions can be drawn from experimental data at relatively low values of the hard scale $Q^2$.

 The experimental data used in our analysis originate from an early Jefferson Lab measurement with a $6$ GeV electron beam. More precise data at higher beam energy and larger $Q^2$ are expected from an ongoing experiment~\cite{Armstrong:2017wfw}, for which the present study already provides phenomenological predictions. A similar analysis for the higher-energy regime of future electron-ion colliders~\cite{AbdulKhalek:2021gbh,Anderle:2021wcy} lies beyond the scope of the present work. In parallel, several complementary exclusive reactions may provide important additional information on the tomography of helium-4. The crossed process, namely timelike Compton scattering in exclusive lepton-pair photoproduction~\cite{Berger:2001xd}, is known~\cite{Grocholski:2019pqj} to be a valuable complement to DVCS, in particular because gluon contributions enter the NLO amplitudes of the two processes in quite different ways~\cite{Pire:2011st,Mueller:2012sma,Moutarde:2013qs}. The more challenging process of double DVCS~\cite{Deja:2023ahc}, i.e. the electroproduction of a lepton pair, for which higher-twist calculations are already available~\cite{Martinez-Fernandez:2025gub}, may also offer new insight into the tomography of helium-4. Because of the charge-conjugation symmetry properties of DVCS and TCS, these reactions probe only the charge-even part of quark GPDs. Access to the charge-odd sector requires considering other exclusive channels, such as diphoton production at large invariant mass~\cite{Pedrak:2017cpp,Pedrak:2020mfm,Grocholski:2022rqj} or photon-plus-vector-meson production~\cite{Boussarie:2016qop,Duplancic:2023kwe}. The theoretical and experimental study of such processes on helium-4 will be necessary to complete the quark tomographic picture of this nucleus. \\

\paragraph*{Acknowledgements.}

We acknowledge useful discussions and correspondence with R.~Dupr\'e, V.~Guzey, and C.~Mezrag.

This research was funded in whole or in part by the National Science Centre, Poland (grant
IMPRESS-U No.~2024/06/Y/ST2/00155 and grant SONATA BIS-15 No.~2025/58/E/ST2/00045). For the purpose of open access, the authors have applied a CC-BY copyright licence to any Author Accepted Manuscript (AAM) version arising from this submission. The research of V.M.F. was funded in part by l’Agence Nationale de la Recherche (ANR), project ANR-23-CE31-0019. 

The data created for the plots shown in this paper are available from the authors upon reasonable request.

\bibliography{bibliography}

@article{Martinez-Fernandez:2026,
    author = "Martinez-Fernandez, V. and Pire, B. and Sznajder, P. and Wagner, J.",
    title = "{Tomography}",
    eprint = "in preparation",
    archivePrefix = "arXiv",
    primaryClass = "hep-ph",
     month = "5",
    year = "2026"
}

@article{Ji:1998xh,
    author = "Ji, Xiang-Dong and Osborne, Jonathan",
    title = "{One loop corrections and all order factorization in deeply virtual Compton scattering}",
    eprint = "hep-ph/9801260",
    archivePrefix = "arXiv",
    reportNumber = "UMD-PP-98-074, DOE-ER-40762-139",
    doi = "10.1103/PhysRevD.58.094018",
    journal = "Phys. Rev. D",
    volume = "58",
    pages = "094018",
    year = "1998"
}

@article{Pedrak:2017cpp,
    author = "Pedrak, A. and Pire, B. and Szymanowski, L. and Wagner, J.",
    title = "{Hard photoproduction of a diphoton with a large invariant mass}",
    eprint = "1708.01043",
    archivePrefix = "arXiv",
    primaryClass = "hep-ph",
    reportNumber = "CPHT-RR-047-082017",
    doi = "10.1103/PhysRevD.96.074008",
    journal = "Phys. Rev. D",
    volume = "96",
    number = "7",
    pages = "074008",
    year = "2017",
    note = "[Erratum: Phys.Rev.D 100, 039901 (2019)]"
}

@article{Pedrak:2020mfm,
    author = "Pedrak, A. and Pire, B. and Szymanowski, L. and Wagner, J.",
    title = "{Electroproduction of a large invariant mass photon pair}",
    eprint = "2003.03263",
    archivePrefix = "arXiv",
    primaryClass = "hep-ph",
    reportNumber = "CPHT-RR010.022020",
    doi = "10.1103/PhysRevD.101.114027",
    journal = "Phys. Rev. D",
    volume = "101",
    number = "11",
    pages = "114027",
    year = "2020"
}

@article{Grocholski:2022rqj,
    author = "Grocholski, Oskar and Pire, Bernard and Sznajder, Pawe{\l} and Szymanowski, Lech and Wagner, Jakub",
    title = "{Phenomenology of diphoton photoproduction at next-to-leading order}",
    eprint = "2204.00396",
    archivePrefix = "arXiv",
    primaryClass = "hep-ph",
    reportNumber = "CPHT-RR020.032022, DESY-22-061",
    doi = "10.1103/PhysRevD.105.094025",
    journal = "Phys. Rev. D",
    volume = "105",
    number = "9",
    pages = "094025",
    year = "2022"
}

@article{Scopetta:2004kj,
    author = "Scopetta, Sergio",
    title = "{Generalized parton distributions of He-3}",
    eprint = "nucl-th/0404014",
    archivePrefix = "arXiv",
    doi = "10.1103/PhysRevC.70.015205",
    journal = "Phys. Rev. C",
    volume = "70",
    pages = "015205",
    year = "2004"
}

@article{Kirch:2005in,
    author = "Kirch, M. and Pobylitsa, P. V. and Goeke, K.",
    title = "{Inequalities for nucleon generalized parton distributions with helicity flip}",
    eprint = "hep-ph/0507048",
    archivePrefix = "arXiv",
    doi = "10.1103/PhysRevD.72.054019",
    journal = "Phys. Rev. D",
    volume = "72",
    pages = "054019",
    year = "2005"
}

@article{Belitsky:2000jk,
    author = "Belitsky, Andrei V. and Mueller, Dieter",
    title = "{Off forward gluonometry}",
    eprint = "hep-ph/0005028",
    archivePrefix = "arXiv",
    doi = "10.1016/S0370-2693(00)00773-5",
    journal = "Phys. Lett. B",
    volume = "486",
    pages = "369--377",
    year = "2000"
}

@article{Guzey:2003jh,
    author = "Guzey, V. and Strikman, M.",
    title = "{DVCS on spinless nuclear targets in impulse approximation}",
    eprint = "hep-ph/0301216",
    archivePrefix = "arXiv",
    reportNumber = "RUB-TP2-19-03",
    doi = "10.1103/PhysRevC.68.015204",
    journal = "Phys. Rev. C",
    volume = "68",
    pages = "015204",
    year = "2003"
}

@article{Kirchner:2003wt,
    author = "Kirchner, A. and Mueller, Dieter",
    title = "{Deeply virtual Compton scattering off nuclei}",
    eprint = "hep-ph/0302007",
    archivePrefix = "arXiv",
    doi = "10.1140/epjc/s2003-01415-x",
    journal = "Eur. Phys. J. C",
    volume = "32",
    pages = "347--375",
    year = "2003"
}

@article{Duplancic:2023kwe,
    author = "Duplan\v{c}i\'c, Goran and Nabeebaccus, Saad and Passek-Kumeri\v{c}ki, Kornelija and Pire, Bernard and Szymanowski, Lech and Wallon, Samuel",
    title = "{Probing chiral-even and chiral-odd leading twist quark generalized parton distributions through the exclusive photoproduction of a \ensuremath{\gamma}\ensuremath{\rho} pair}",
    eprint = "2302.12026",
    archivePrefix = "arXiv",
    primaryClass = "hep-ph",
    doi = "10.1103/PhysRevD.107.094023",
    journal = "Phys. Rev. D",
    volume = "107",
    number = "9",
    pages = "094023",
    year = "2023"
}

@article{Boussarie:2016qop,
    author = "Boussarie, R. and Pire, B. and Szymanowski, L. and Wallon, S.",
    title = "{Exclusive photoproduction of a $\gamma\,\rho$ pair with a large invariant mass}",
    eprint = "1609.03830",
    archivePrefix = "arXiv",
    primaryClass = "hep-ph",
    reportNumber = "LPT-ORSAY-16-58, CPHT-RR038.072016, LPT-Orsay-16-58",
    doi = "10.1007/JHEP02(2017)054",
    journal = "JHEP",
    volume = "02",
    pages = "054",
    year = "2017",
    note = "[Erratum: JHEP 10, 029 (2018)]"
}

@article{Ralston:2001xs,
    author = "Ralston, John P. and Pire, Bernard",
    title = "{Femtophotography of protons to nuclei with deeply virtual Compton scattering}",
    eprint = "hep-ph/0110075",
    archivePrefix = "arXiv",
    reportNumber = "CPHT-S038-0901",
    doi = "10.1103/PhysRevD.66.111501",
    journal = "Phys. Rev. D",
    volume = "66",
    pages = "111501",
    year = "2002"
}

@article{Pire:2011st,
    author = "Pire, B. and Szymanowski, L. and Wagner, J.",
    title = "{NLO corrections to timelike, spacelike and double deeply virtual Compton scattering}",
    eprint = "1101.0555",
    archivePrefix = "arXiv",
    primaryClass = "hep-ph",
    doi = "10.1103/PhysRevD.83.034009",
    journal = "Phys. Rev. D",
    volume = "83",
    pages = "034009",
    year = "2011"
}

@article{Cano:2003ju,
    author = "Cano, F. and Pire, B.",
    title = "{Deep electroproduction of photons and mesons on the deuteron}",
    eprint = "hep-ph/0307231",
    archivePrefix = "arXiv",
    doi = "10.1140/epja/i2003-10127-x",
    journal = "Eur. Phys. J. A",
    volume = "19",
    pages = "423--438",
    year = "2004"
}

@article{Berger:2001zb,
    author = "Berger, Edgar R. and Cano, F. and Diehl, M. and Pire, B.",
    title = "{Generalized parton distributions in the deuteron}",
    eprint = "hep-ph/0106192",
    archivePrefix = "arXiv",
    reportNumber = "CPHT-S029-0601, DESY-01-082",
    doi = "10.1103/PhysRevLett.87.142302",
    journal = "Phys. Rev. Lett.",
    volume = "87",
    pages = "142302",
    year = "2001"
}

@article{Moutarde:2013qs,
    author = "Moutarde, H. and Pire, B. and Sabatie, F. and Szymanowski, L. and Wagner, J.",
    title = "{Timelike and spacelike deeply virtual Compton scattering at next-to-leading order}",
    eprint = "1301.3819",
    archivePrefix = "arXiv",
    primaryClass = "hep-ph",
    reportNumber = "CPHT-RR002.0113, IRFU-13-01",
    doi = "10.1103/PhysRevD.87.054029",
    journal = "Phys. Rev. D",
    volume = "87",
    number = "5",
    pages = "054029",
    year = "2013"
}

@article{Diehl:2002he,
    author = "Diehl, M.",
    title = "{Generalized parton distributions in impact parameter space}",
    eprint = "hep-ph/0205208",
    archivePrefix = "arXiv",
    doi = "10.1007/s10052-002-1016-9",
    journal = "Eur. Phys. J. C",
    volume = "25",
    pages = "223--232",
    year = "2002",
    note = "[Erratum: Eur.Phys.J.C 31, 277--278 (2003)]"
}

@article{Pire:1998nw,
    author = "Pire, B. and Soffer, Jacques and Teryaev, O.",
    title = "{Positivity constraints for off - forward parton distributions}",
    eprint = "hep-ph/9804284",
    archivePrefix = "arXiv",
    reportNumber = "CPT-98-P-3633",
    doi = "10.1007/s100529901063",
    journal = "Eur. Phys. J. C",
    volume = "8",
    pages = "103--106",
    year = "1999"
}

@article{Belitsky:2001ns,
    author = "Belitsky, Andrei V. and Mueller, Dieter and Kirchner, A.",
    title = "{Theory of deeply virtual Compton scattering on the nucleon}",
    eprint = "hep-ph/0112108",
    archivePrefix = "arXiv",
    reportNumber = "DOE-ER-40762-009, UMD-PP-02-011, YITP-SB-01-51",
    doi = "10.1016/S0550-3213(02)00144-X",
    journal = "Nucl. Phys. B",
    volume = "629",
    pages = "323--392",
    year = "2002"
}

@article{Bertone:2017gds,
    author = "Bertone, Valerio",
    editor = "Klein, Uta",
    title = "{APFEL++: A new PDF evolution library in C++}",
    eprint = "1708.00911",
    archivePrefix = "arXiv",
    primaryClass = "hep-ph",
    doi = "10.22323/1.297.0201",
    journal = "PoS",
    volume = "DIS2017",
    pages = "201",
    year = "2018"
}

@article{Deja:2023ahc,
    author = "Deja, K. and Martinez-Fernandez, V. and Pire, B. and Sznajder, P. and Wagner, J.",
    title = "{Phenomenology of double deeply virtual Compton scattering in the era of new experiments}",
    eprint = "2303.13668",
    archivePrefix = "arXiv",
    primaryClass = "hep-ph",
    reportNumber = "CPHT-RR012.032022",
    doi = "10.1103/PhysRevD.107.094035",
    journal = "Phys. Rev. D",
    volume = "107",
    number = "9",
    pages = "094035",
    year = "2023"
}

@article{Braun:2012bg,
    author = "Braun, V. M. and Manashov, A. N. and Pirnay, B.",
    title = "{Finite-t and target mass corrections to DVCS on a scalar target}",
    eprint = "hep-ph/1205.3332",
    archivePrefix = "arXiv",
    primaryClass = "hep-ph",
    reportNumber = "IPHT-T12-038",
    doi = "10.1103/PhysRevD.86.014003",
    journal = "Phys. Rev. D",
    volume = "86",
    pages = "014003",
    year = "2012"
}

@article{Braun:2022qly,
    author = "Braun, V. M. and Ji, Yao and Manashov, A. N.",
    title = "{Next-to-leading-power kinematic corrections to DVCS: a scalar target}",
    eprint = "hep-ph/2211.04902",
    archivePrefix = "arXiv",
    primaryClass = "hep-ph",
    reportNumber = "TUM-HEP-1432/22, DESY-22-169",
    doi = "10.1007/JHEP01(2023)078",
    journal = "JHEP",
    volume = "01",
    pages = "078",
    year = "2023"
}

@article{Polyakov:1999gs,
    author = "Polyakov, Maxim V. and Weiss, C.",
    title = "{Skewed and double distributions in pion and nucleon}",
    eprint = "hep-ph/9902451",
    archivePrefix = "arXiv",
    reportNumber = "RUB-TPII-1-99",
    doi = "10.1103/PhysRevD.60.114017",
    journal = "Phys. Rev. D",
    volume = "60",
    pages = "114017",
    year = "1999"
}

@article{Braun:2014sta,
    author = {Braun, Vladimir M. and Manashov, Alexander N. and M\"uller, Dieter and Pirnay, Bjoern M.},
    title = "{Deeply Virtual Compton Scattering to the twist-four accuracy: Impact of finite-$t$ and target mass corrections}",
    eprint = "1401.7621",
    archivePrefix = "arXiv",
    primaryClass = "hep-ph",
    doi = "10.1103/PhysRevD.89.074022",
    journal = "Phys. Rev. D",
    volume = "89",
    number = "7",
    pages = "074022",
    year = "2014"
}

@article{Burkardt:2002hr,
    author = "Burkardt, Matthias",
    title = "{Impact parameter space interpretation for generalized parton distributions}",
    eprint = "0207047",
    primaryClass = "hep-ph",
    archivePrefix = "arXiv",
    doi = "10.1142/S0217751X03012370",
    journal = "Int. J. Mod. Phys. A",
    volume = "18",
    pages = "173--208",
    year = "2003"
}

@article{Berger:2001xd,
      author         = "Berger, Edgar R. and Diehl, M. and Pire, B.",
      title          = "{Time - like Compton scattering: Exclusive
                        photoproduction of lepton pairs}",
      journal        = "Eur. Phys. J.",
      volume         = "C23",
      year           = "2002",
      pages          = "675-689",
      doi            = "10.1007/s100520200917",
      eprint         = "hep-ph/0110062",
      archivePrefix  = "arXiv",
      primaryClass   = "hep-ph",
      reportNumber   = "CPHT-S010-0201, DESY-01-119",
      SLACcitation   = "%%CITATION = HEP-PH/0110062;%%"
}

@article{Belitsky:2005qn,
      author         = "Belitsky, A. V. and Radyushkin, A. V.",
      title          = "{Unraveling hadron structure with generalized parton
                        distributions}",
      journal        = "Phys. Rept.",
      volume         = "418",
      year           = "2005",
      pages          = "1-387",
      doi            = "10.1016/j.physrep.2005.06.002",
      eprint         = "hep-ph/0504030",
      archivePrefix  = "arXiv",
      primaryClass   = "hep-ph",
      reportNumber   = "JLAB-THY-04-34",
      SLACcitation   = "%%CITATION = HEP-PH/0504030;%%"
}

@article{AbdulKhalek:2021gbh,
    author = "Abdul Khalek, R. and others",
    title = "{Science Requirements and Detector Concepts for the Electron-Ion Collider}: {EIC Yellow Report}",
    eprint = "physics.ins-det/2103.05419",
    archivePrefix = "arXiv",
    primaryClass = "physics.ins-det",
    reportNumber = "BNL-220990-2021-FORE, JLAB-PHY-21-3198, LA-UR-21-20953",
    doi = "10.1016/j.nuclphysa.2022.122447",
    journal = "Nucl. Phys. A",
    volume = "1026",
    pages = "122447",
    year = "2022"
}

@article{Anderle:2021wcy,
    author = "Anderle, Daniele P. and others",
    title = "{Electron-ion collider in China}",
    eprint = "nucl-ex/2102.09222",
    archivePrefix = "arXiv",
    primaryClass = "nucl-ex",
    doi = "10.1007/s11467-021-1062-0",
    journal = "Front. Phys. (Beijing)",
    volume = "16",
    number = "6",
    pages = "64701",
    year = "2021"
}

@article{Grocholski:2019pqj,
    author = "Grocholski, O. and Moutarde, H. and Pire, B. and Sznajder, P. and Wagner, J.",
    title = "{Data-driven study of timelike Compton scattering}",
    eprint = "1912.09853",
    archivePrefix = "arXiv",
    primaryClass = "hep-ph",
    doi = "10.1140/epjc/s10052-020-7700-9",
    journal = "Eur. Phys. J. C",
    volume = "80",
    number = "2",
    pages = "171",
    year = "2020"
}

@article{Mueller:2012sma,
    author = "Mueller, Dieter and Pire, B. and Szymanowski, L. and Wagner, J.",
    title = "{On timelike and spacelike hard exclusive reactions}",
    eprint = "1203.4392",
    archivePrefix = "arXiv",
    primaryClass = "hep-ph",
    reportNumber = "CPHT-RR015.0312",
    doi = "10.1103/PhysRevD.86.031502",
    journal = "Phys. Rev. D",
    volume = "86",
    pages = "031502",
    year = "2012"
}

@article{Fucini_2018,
   title={Coherent deeply virtual Compton scattering off 
<mml:math xmlns:mml="http://www.w3.org/1998/Math/MathML"><mml:mmultiscripts><mml:mi>He</mml:mi><mml:mprescripts /><mml:none /><mml:mn>4</mml:mn></mml:mmultiscripts></mml:math>},
   volume={98},
   ISSN={2469-9993},
   url={http://dx.doi.org/10.1103/PhysRevC.98.015203},
   DOI={10.1103/physrevc.98.015203},
   number={1},
   journal={Physical Review C},
   publisher={American Physical Society (APS)},
   author={Fucini, Sara and Scopetta, Sergio and Viviani, Michele},
   year={2018},
   month=jul }

@article{CLAS:2017udk,
    author = "Hattawy, M. and others",
    collaboration = "CLAS",
    title = "{First Exclusive Measurement of Deeply Virtual Compton Scattering off $^4$He: Toward the 3D Tomography of Nuclei}",
    eprint = "1707.03361",
    archivePrefix = "arXiv",
    primaryClass = "nucl-ex",
    reportNumber = "JLAB-PHY-17-2543",
    doi = "10.1103/PhysRevLett.119.202004",
    journal = "Phys. Rev. Lett.",
    volume = "119",
    number = "20",
    pages = "202004",
    year = "2017"
}

@article{Martinez-Fernandez:2025gub,
    author = "Martinez-Fernandez, V. and Pire, B. and Sznajder, P. and Wagner, J.",
    title = "{Timelike Compton scattering on a spin-0 target with kinematic twist-4 precision}",
    eprint = "2503.02461",
    archivePrefix = "arXiv",
    primaryClass = "hep-ph",
    reportNumber = "CPHT-RR088.122024",
    doi = "10.1103/PhysRevD.111.074034",
    journal = "Phys. Rev. D",
    volume = "111",
    number = "7",
    pages = "074034",
    year = "2025"
}

@article{Diehl:2003ny,
	author = "Diehl, M.",
	title = "{Generalized parton distributions}",
	eprint = "hep-ph/0307382",
	archivePrefix = "arXiv",
	reportNumber = "DESY-THESIS-2003-018",
	doi = "10.1016/j.physrep.2003.08.002",
	journal = "Phys. Rept.",
	volume = "388",
	pages = "41--277",
	year = "2003"
}

@article{Bacchetta:2004jz,
	author = "Bacchetta, Alessandro and D'Alesio, Umberto and Diehl, Markus and Miller, C. Andy",
	title = "{Single-spin asymmetries: The Trento conventions}",
	eprint = "hep-ph/0410050",
	archivePrefix = "arXiv",
	reportNumber = "DESY-04-193",
	doi = "10.1103/PhysRevD.70.117504",
	journal = "Phys. Rev. D",
	volume = "70",
	pages = "117504",
	year = "2004"
}

@article{Belitsky:2002tf,
	author = "Belitsky, Andrei V. and Mueller, Dieter",
	title = "{Exclusive electroproduction of lepton pairs as a probe of nucleon structure}",
	eprint = "hep-ph/0210313",
	archivePrefix = "arXiv",
	reportNumber = "DOE-ER-40762-270, UMD-PP-03-024",
	doi = "10.1103/PhysRevLett.90.022001",
	journal = "Phys. Rev. Lett.",
	volume = "90",
	pages = "022001",
	year = "2003"
}

@article{Martinez-Fernandez:2025rcg,
	author = "Mart{\'\i}nez-Fern{\'a}ndez, V{\'\i}ctor and Mezrag, C{\'e}dric",
	title = "{Dispersion relations of deeply virtual Compton scattering: investigating twist-4 kinematic power corrections}",
	eprint = "2509.05059",
	archivePrefix = "arXiv",
	primaryClass = "hep-ph",
	month = "9",
	year = "2025"
}

@article{Martinez-Fernandez:2025jvk,
	author = "Mart{\'\i}nez-Fern{\'a}ndez, V{\'\i}ctor and Binosi, Daniele and Mezrag, C{\'e}dric and Yao, Zhao-Qian",
	title = "{Constraining the Energy Momentum Tensor through DVCS Dispersion Relation beyond Leading Power}",
	eprint = "2509.06669",
	archivePrefix = "arXiv",
	primaryClass = "hep-ph",
	month = "9",
	year = "2025"
}

@article{PhysRevLett.78.610,
	title = {Gauge-Invariant Decomposition of Nucleon Spin},
	author = {Ji, Xiangdong},
	journal = {Phys. Rev. Lett.},
	volume = {78},
	issue = {4},
	pages = {610--613},
	numpages = {0},
	year = {1997},
	month = {Jan},
	publisher = {American Physical Society},
	doi = {10.1103/PhysRevLett.78.610},
	url = {https://link.aps.org/doi/10.1103/PhysRevLett.78.610}
}

@article{Lorce:2018egm,
	author = "Lorc{\'e}, C{\'e}dric and Moutarde, Herv{\'e} and Trawi{\'n}ski, Arkadiusz P.",
	title = "{Revisiting the mechanical properties of the nucleon}",
	eprint = "1810.09837",
	archivePrefix = "arXiv",
	primaryClass = "hep-ph",
	doi = "10.1140/epjc/s10052-019-6572-3",
	journal = "Eur. Phys. J. C",
	volume = "79",
	number = "1",
	pages = "89",
	year = "2019"
}

@article{Braun:2025xlp,
	author = "Braun, V. M. and Ji, Yao and Manashov, A. N.",
	title = "{Kinematic power corrections to deeply virtual Compton scattering to twist-six accuracy}",
	eprint = "2501.08185",
	archivePrefix = "arXiv",
	primaryClass = "hep-ph",
	reportNumber = "TUM-HEP-1551/25, DESY 24--223",
	doi = "10.1103/PhysRevD.111.076011",
	journal = "Phys. Rev. D",
	volume = "111",
	number = "7",
	pages = "076011",
	year = "2025"
}

@article{KLEISS198461,
	title = {The cross section for e+e $\to$ e+e-e+e-},
	journal = {Nuclear Physics B},
	volume = {241},
	number = {1},
	pages = {61-74},
	year = {1984},
	issn = {0550-3213},
	doi = {https://doi.org/10.1016/0550-3213(84)90197-4},
	author = {Ronald Kleiss}
}

@article{KLEISS1985235,
	title = {Spinor techniques for calculating pp $\to$ W+-/Z0 + jets},
	journal = {Nuclear Physics B},
	volume = {262},
	number = {2},
	pages = {235-262},
	year = {1985},
	issn = {0550-3213},
	doi = {https://doi.org/10.1016/0550-3213(85)90285-8},
	author = {R. Kleiss and W.J. Stirling}
}

@article{PhysRevLett.119.202004,
	title = {First Exclusive Measurement of Deeply Virtual Compton Scattering off $^{4}\mathrm{He}$: Toward the 3D Tomography of Nuclei},
	author = {Hattawy, M. and Baltzell, N. A. and Dupr\'e, R. and Hafidi, K. and Stepanyan, S. and B\"ultmann, S. and De Vita, R. and El Alaoui, A. and El Fassi, L. and Egiyan, H. and Girod, F. X. and Guidal, M. and Jenkins, D. and Liuti, S. and Perrin, Y. and Torayev, B. and Voutier, E. and Adhikari, K. P. and Adhikari, S. and Adikaram, D. and Akbar, Z. and Amaryan, M. J. and Anefalos Pereira, S. and Armstrong, Whitney R. and Avakian, H. and Ball, J. and Bashkanov, M. and Battaglieri, M. and Batourine, V. and Bedlinskiy, I. and Biselli, A. S. and Boiarinov, S. and Briscoe, W. J. and Brooks, W. K. and Burkert, V. D. and Thanh Cao, Frank and Carman, D. S. and Celentano, A. and Charles, G. and Chetry, T. and Ciullo, G. and Clark, L. and Colaneri, L. and Cole, P. L. and Contalbrigo, M. and Cortes, O. and Crede, V. and D'Angelo, A. and Dashyan, N. and De Sanctis, E. and Deur, A. and Djalali, C. and Elouadrhiri, L. and Eugenio, P. and Fedotov, G. and Fegan, S. and Fersch, R. and Filippi, A. and Fleming, J. A. and Forest, T. A. and Fradi, A. and Gar\ifmmode \mbox{\c{c}}\else \c{c}\fi{}on, M. and Gevorgyan, N. and Ghandilyan, Y. and Gilfoyle, G. P. and Giovanetti, K. L. and Gleason, C. and Gohn, W. and Golovatch, E. and Gothe, R. W. and Griffioen, K. A. and Guo, L. and Hakobyan, H. and Hanretty, C. and Harrison, N. and Heddle, D. and Hicks, K. and Holtrop, M. and Hughes, S. M. and Ireland, D. G. and Ishkhanov, B. S. and Isupov, E. L. and Jiang, H. and Joo, K. and Joosten, S. and Keller, D. and Khachatryan, G. and Khachatryan, M. and Khandaker, M. and Kim, A. and Kim, W. and Klein, A. and Klein, F. J. and Kubarovsky, V. and Kuhn, S. E. and Kuleshov, S. V. and Lanza, L. and Lenisa, P. and Livingston, K. and Lu, H. Y. and MacGregor, I. J. D. and Markov, N. and Mayer, M. and McCracken, M. E. and McKinnon, B. and Meyer, C. A. and Meziani, Z. E. and Mineeva, T. and Mirazita, M. and Mokeev, V. and Montgomery, R. A. and Moutarde, H. and Movsisyan, A. and Munoz Camacho, C. and Nadel-Turonski, P. and Net, L. A. and Niccolai, S. and Niculescu, G. and Niculescu, I. and Osipenko, M. and Ostrovidov, A. I. and Paolone, M. and Paremuzyan, R. and Park, K. and Pasyuk, E. and Phelps, E. and Phelps, W. and Pisano, S. and Pogorelko, O. and Price, J. W. and Prok, Y. and Protopopescu, D. and Ripani, M. and Ritchie, B. G. and Rizzo, A. and Rosner, G. and Rossi, P. and Sabati\'e, F. and Salgado, C. and Schumacher, R. A. and Seder, E. and Sharabian, Y. G. and Simonyan, A. and Skorodumina, Iu. and Smith, G. D. and Sokhan, D. and Sparveris, N. and Strauch, S. and Taiuti, M. and Ungaro, M. and Voskanyan, H. and Walford, N. K. and Watts, D. P. and Wei, X. and Weinstein, L. B. and Wood, M. H. and Zachariou, N. and Zana, L. and Zhang, J. and Zhao, Z. W.},
	collaboration = {CLAS Collaboration},
	journal = {Phys. Rev. Lett.},
	volume = {119},
	issue = {20},
	pages = {202004},
	numpages = {7},
	year = {2017},
	month = {Nov},
	publisher = {American Physical Society},
	doi = {10.1103/PhysRevLett.119.202004},
	url = {https://link.aps.org/doi/10.1103/PhysRevLett.119.202004}
}

@article{PhysRevC.104.025203,
	title = {Measurement of deeply virtual Compton scattering off $^{4}\mathrm{He}$ with the CEBAF Large Acceptance Spectrometer at Jefferson Lab},
	author = {Dupr\'e, R. and Hattawy, M. and Baltzell, N. A. and B\"ultmann, S. and De Vita, R. and El Alaoui, A. and El Fassi, L. and Egiyan, H. and Girod, F. X. and Guidal, M. and Hafidi, K. and Jenkins, D. and Liuti, S. and Perrin, Y. and Stepanyan, S. and Torayev, B. and Voutier, E. and Amaryan, M. J. and Armstrong, W. R. and Atac, H. and Ayerbe Gayoso, C. and Barion, L. and Battaglieri, M. and Bedlinskiy, I. and Benmokhtar, F. and Bianconi, A. and Biselli, A. S. and Bondi, M. and Boss\`u, F. and Boiarinov, S. and Briscoe, W. J. and Bulumulla, D. and Burkert, V. and Carman, D. S. and Carvajal, J. C. and Caudron, M. and Celentano, A. and Chatagnon, P. and Chesnokov, V. and Chetry, T. and Ciullo, G. and Clary, B. A. and Cole, P. L. and Contalbrigo, M. and Costantini, G. and Crede, V. and D'Angelo, A. and Dashyan, N. and Defurne, M. and Deur, A. and Diehl, S. and Djalali, C. and Ehrhart, M. and Elouadrhiri, L. and Eugenio, P. and Fegan, S. and Filippi, A. and Forest, T. A. and Ghandilyan, Y. and Gilfoyle, G. P. and Gothe, R. W. and Griffioen, K. A. and Hakobyan, H. and Hayward, T. B. and Hicks, K. and Hobart, A. and Holtrop, M. and Ilieva, Y. and Ireland, D. G. and Isupov, E. L. and Jo, H. S. and Joo, K. and Joosten, S. and Keller, D. and Khachatryan, G. and Khanal, A. and Khandaker, M. and Kim, A. and Kim, W. and Kripko, A. and Kubarovsky, V. and Kuhn, S. E. and Lanza, L. and Livingston, K. and Kabir, M. L. and Leali, M. and Lenisa, P. and MacGregor, I. J. D. and Marchand, D. and Markov, N. and Mascagna, V. and Mayer, M. and McKinnon, B. and Mirazita, M. and Mokeev, V. I. and Neupane, K. and Niccolai, S. and O'Connell, T. R. and Osipenko, M. and Paolone, M. and Pappalardo, L. L. and Paremuzyan, R. and Pasyuk, E. and Payette, D. and Phelps, W. and Pivnyuk, N. and Pogorelko, O. and Poudel, J. and Prok, Y. and Ripani, M. and Ritman, J. and Rizzo, A. and Rosner, G. and Rossi, P. and Rowley, J. and Sabati\'e, F. and Salgado, C. and Schmidt, A. and Schumacher, R. and Sergeyeva, V. and Sharabian, Y. and Shrestha, U. and Sokhan, D. and Soto, O. and Sparveris, N. and Strakovsky, I. I. and Strauch, S. and Tyler, N. and Ungaro, M. and Venturelli, L. and Voskanyan, H. and Vossen, A. and Watts, D. and Wei, K. and Wei, X. and Weinstein, L. B. and Wishart, R. and Wood, M. H. and Yale, B. and Zachariou, N. and Zhang, J.},
	collaboration = {CLAS Collaboration},
	journal = {Phys. Rev. C},
	volume = {104},
	issue = {2},
	pages = {025203},
	numpages = {25},
	year = {2021},
	month = {Aug},
	publisher = {American Physical Society},
	doi = {10.1103/PhysRevC.104.025203},
	url = {https://link.aps.org/doi/10.1103/PhysRevC.104.025203}
}

@article{Goloskokov:2006hr,
    author = "Goloskokov, S. V. and Kroll, P.",
    title = "{The Longitudinal cross-section of vector meson electroproduction}",
    eprint = "hep-ph/0611290",
    archivePrefix = "arXiv",
    reportNumber = "WU-B-06-02, WU B 06-02",
    doi = "10.1140/epjc/s10052-007-0228-4",
    journal = "Eur. Phys. J. C",
    volume = "50",
    pages = "829--842",
    year = "2007"
}

@article{JeffersonLabHallA:2013cus,
    author = "Camsonne, A. and others",
    collaboration = "Jefferson Lab Hall A",
    title = "{JLab Measurement of the $^4$He Charge Form Factor at Large Momentum Transfers}",
    eprint = "1309.5297",
    archivePrefix = "arXiv",
    primaryClass = "nucl-ex",
    reportNumber = "JLAB-PHY-13-1798",
    doi = "10.1103/PhysRevLett.112.132503",
    journal = "Phys. Rev. Lett.",
    volume = "112",
    number = "13",
    pages = "132503",
    year = "2014"
}

@article{Arnold:1978qs,
    author = "Arnold, R. G. and others",
    title = "{Elastic electron Scattering from He-3 and He-4 at High Momentum Transfer}",
    reportNumber = "SLAC-PUB-2093",
    doi = "10.1103/PhysRevLett.40.1429",
    journal = "Phys. Rev. Lett.",
    volume = "40",
    pages = "1429",
    year = "1978"
}

@article{Frosch:1967pz,
    author = "Frosch, R. F. and McCarthy, J. S. and Rand, R. e. and Yearian, M. R.",
    title = "{Structure of the He-4 nucleus from elastic electron scattering}",
    doi = "10.1103/PhysRev.160.874",
    journal = "Phys. Rev.",
    volume = "160",
    pages = "874--879",
    year = "1967"
}

@article{Ottermann:1985km,
    author = "Ottermann, C. R. and Kobschall, G. and Maurer, K. and Rohrich, K. and Schmitt, C. and Walther, V. H.",
    title = "{ELASTIC ELECTRON SCATTERING FROM HE-3 AND HE-4}",
    doi = "10.1016/0375-9474(85)90554-8",
    journal = "Nucl. Phys. A",
    volume = "436",
    pages = "688--698",
    year = "1985"
}

@article{Repellin:1965vba,
    author = "Repellin, J. P. and Lehmann, P. and Lefran{\c{c}}ois, J. and Isabelle, D. B.",
    title = "{Elastic electron scattering on helium 4}",
    doi = "10.1016/0031-9163(65)90172-1",
    journal = "Phys. Lett.",
    volume = "16",
    number = "2",
    pages = "169--170",
    year = "1965"
}

@article{AbdulKhalek:2022fyi,
    author = "Abdul Khalek, Rabah and Gauld, Rhorry and Giani, Tommaso and Nocera, Emanuele R. and Rabemananjara, Tanjona R. and Rojo, Juan",
    title = "{nNNPDF3.0: evidence for a modified partonic structure in heavy nuclei}",
    eprint = "2201.12363",
    archivePrefix = "arXiv",
    primaryClass = "hep-ph",
    reportNumber = "Nikhef-2021-028, BONN-TH-2021-14",
    doi = "10.1140/epjc/s10052-022-10417-7",
    journal = "Eur. Phys. J. C",
    volume = "82",
    number = "6",
    pages = "507",
    year = "2022"
}

@misc{Armstrong:2017wfw,
    author = "Armstrong, Whitney and others",
    title = "{Partonic Structure of Light Nuclei}",
    eprint = "1708.00888",
    archivePrefix = "arXiv",
    primaryClass = "nucl-ex",
    month = "8",
    year = "2017"
}

@article{Bertone:2022frx,
    author = "Bertone, Valerio and Dutrieux, Herv{\'e} and Mezrag, C{\'e}dric and Morgado, Jos{\'e} M. and Moutarde, Herv{\'e}",
    title = "{Revisiting evolution equations for generalised parton distributions}",
    eprint = "2206.01412",
    archivePrefix = "arXiv",
    primaryClass = "hep-ph",
    doi = "10.1140/epjc/s10052-022-10793-0",
    journal = "Eur. Phys. J. C",
    volume = "82",
    number = "10",
    pages = "888",
    year = "2022"
}

@article{Radyushkin:1998es,
    author = "Radyushkin, A. V.",
    title = "{Double distributions and evolution equations}",
    eprint = "hep-ph/9805342",
    archivePrefix = "arXiv",
    reportNumber = "JLAB-THY-98-16",
    doi = "10.1103/PhysRevD.59.014030",
    journal = "Phys. Rev. D",
    volume = "59",
    pages = "014030",
    year = "1999"
}

\appendix


\section{Kinematics}\label{sec::kinematics}

In this section, we describe the kinematics of the process~(\ref{reaction}). The core of our calculation is performed in the ``target rest frame'' (TRF) that is related to the usual Trento frame~\cite{Bacchetta:2004jz} by a 180\textdegree\ rotation around the $y$-axis. This TRF frame coincides with TRF-I of the DDVCS studies in Refs.~\cite{Belitsky:2002tf,Deja:2023ahc} and it is depicted in Fig.~\ref{fig::TRFandTrento}, allowing us to stay consistent with previous literature on exclusive processes.

\subsection{Momenta parameterization}

In TRF, the target is at rest so that the initial-state momentum is given by
\begin{equation}
	p^\mu = (M, 0,0,0)\,,
\end{equation}
while the $z$-axis is opposite to the incoming photon momentum,
\begin{equation}
	q^\mu = (q^0, 0,0,-|q^3|) \,.
\end{equation}
Accounting for its spacelike virtuality, $q^2=-Q^2 < 0$, as well as the {\it nuclear Bjorken variable,} $x_A=Q^2/(2pq) \underset{\rm TRF}{=} Q^2/(2Mq^0)$, we find
\begin{equation}
	q^\mu = \frac{Q}{\omega}(1,0,0,-\sqrt{1+\omega^2}) \,, \quad \omega = \frac{2x_A M}{Q}\,.
\end{equation}

The final-state nucleus has a momentum
\begin{equation}
	p'^\mu = (p'^0, |\vec{p}^{\,\prime}|s_N c_\varphi, |\vec{p}^{\,\prime}|s_N s_\varphi, |\vec{p}^{\,\prime}|c_N)\,,
\end{equation}
where $s_N = \sin\theta_N$, $c_N = \cos\theta_N$ for $\theta_N$ the polar angle with respect to TRF, and equivalently $s_\varphi = \sin\varphi$, $c_\varphi=\cos\varphi$ for its azimuth $\varphi$. With definition of the Mandelstam $t=(p'-p)^2$ and the above parameterization of the initial- and final-state momenta, we find
\begin{equation}
	p'^0 = M-\frac{t}{2M}\,,\quad |\vec{p}^{\,\prime}| = \sqrt{-t\left( 1-\frac{t}{4M^2} \right)}\,.
\end{equation}
Momentum conservation fixes the angle $\theta_N$. Indeed, the momentum of the the real outgoing photon is related to the momentum of incoming photon and the momentum transfer ($\Delta = p'-p$) by
\begin{equation}\label{qPrime2}
	q' = q - \Delta \Rightarrow q'^2 = 0 = -Q^2 + t - 2q\Delta\,,
\end{equation}
where
\begin{equation}
	\Delta^\mu = (p'-p)^\mu = \left( -\frac{t}{2M} , \vec{p}^{\,\prime} \right) \Rightarrow q\Delta = \frac{Q}{\omega}\left( -\frac{t}{2M} + \sqrt{1+\omega^2}|\vec{p}^{\,\prime}|c_N \right) \,.
\end{equation}
Introducing this product into Eq.~(\ref{qPrime2}), we can solve $c_N$ to be
\begin{equation}
	c_N = -\frac{\omega^2(Q^2-t)-2x_At}{4Mx_A\sqrt{1+\omega^2}|\vec{p}^{\,\prime}|} \,.
\end{equation}

Thanks to the null virtuality of the outgoing photon, $q'^\mu = (q'^0,\vec{q}^{\,\prime})$ such that $(q'^0)^2 = |\vec{q}^{\,\prime}|^2$. Again, due to momentum conservation
\begin{equation}
	t = (p'-p)(q-q') = p'\Delta - pq + pq'\,.
\end{equation}  
With the above parameterization,
\begin{equation}
	\left.
		\begin{aligned}
			p'\Delta & = \underbrace{\bp\Delta}_{0} - p\Delta = -M\Delta^0 = \frac{t}{2}\,,\\
			pq & = p^0 q^0 = \frac{MQ}{\omega} \,,\\
			pq' & = p^0 q'^0 = Mq'^0\,,
		\end{aligned}
	\right\} \Rightarrow q'^0 = \frac{t}{2M} + \frac{Q}{\omega} \,,
\end{equation}
which leads us to
\begin{equation}
	q'^\mu = \left( \frac{t}{2M} + \frac{Q}{\omega} , -\vec{p}^{\,\prime} + \vec{q} \right) \,.
\end{equation}
In TRF and for a massless lepton beam,
\begin{equation}
	k^\mu = E(1,\sin\theta_e,0,\cos\theta_e)\,,\quad E = \frac{Q}{y\omega}\,,
\end{equation}
where the inelasticity variable has been defined as $y=pq/(pk) \underset{\rm TRF}{=} Q/(\omega E)$. 

Finally, the momentum of the scattered electron reads
\begin{equation}
	k' = -\Delta + k - q' \Rightarrow k'^\mu = \left( \frac{(1-y)Q}{y\omega}, \vec{k} - \vec{q} \right)\,. 
\end{equation}

Note that only the momenta $p', q'$ have a dependence on the azimuthal angle $\varphi$ whose relation to its Trento counterpart $\phi$ is
\begin{equation}
	\varphi = \left\{
		\begin{aligned}
			& \pi-\phi\,, \mbox{ if } \phi\in [0,\pi]\,,\\
			& 3\pi - \phi\,, \mbox{ if } \phi\in (\pi,2\pi)\,.
		\end{aligned}
	\right.
\end{equation}

\subsection{Longitudinal vs transverse plane}

The collinear factorization that gives rise to GPDs and CFFs requires a separation of dominant longitudinal kinematics (characterized by lightlike vectors) and subleading transverse components that are orthogonal to the former. In such a framework, any four-vector can be written as
\begin{equation}
	v^\mu = v^+ n'^\mu + v^-n^\mu + v_\perp^\mu\,,
\end{equation}
where $n^2=n'^2=0$, $v_\perp n = v_\perp n' = 0$ and $nn'\neq 0$. Following the procedure of Refs.~\cite{Braun:2012bg,Martinez-Fernandez:2025gub} and to obtain a leading-twist limit for the Compton tensor that is compatible with charge conservation ($q_\nu T^{\mu\nu}_{\rm LT}=T^{\mu\nu}_{\rm LT}q'_\mu=0$), we choose purely longitudinal photon momenta:
\begin{equation}
	n^\mu = \alpha q^\mu + \beta q'^\mu\,,\quad n'^\mu = \alpha' q^\mu + \beta' q'^\mu\,,
\end{equation}
where $\alpha$, $\beta$, $\alpha'$, $\beta'$ are real parameters constrained by $n^2 = n'^2 = 0$, together with the normalization
\begin{equation}\label{nn'}
	nn' = -\Delta q' = \frac{t + Q^2}{2} \neq 0\,.
\end{equation}
The solution after rescaling $Q^2 n\to n$ and $n'/Q^2\to n'$ is
\begin{align}
	n^\mu & = Q^2 q'^\mu \,, \label{n} \\
	n'^\mu & = -\frac{1}{Q^2} q^\mu + \frac{1}{Q^2+t}q'^\mu \label{nPrime} \,.
\end{align}
Alternatively,
\begin{align}
	q^\mu & = \frac{1}{Q^2+t}n^\mu - Q^2 n'^\mu \,,\\
	q'^\mu & = \frac{1}{Q^2}n^\mu \,.
\end{align}
An immediate consequence is the vanishing of the transverse component of the momentum transfer, this is $\Delta_\perp = (q-q')_\perp=0$, while $p_\perp=p'_\perp=\bp_\perp$ is the only momentum with transverse components. Taking into account that $\Delta\bp = 0$, the definition of the skewness $\xi = -\Delta n/(2\bp n)$ and the square $\bp^{\,2} = M^2 - t/4$, we find:
\begin{equation}\label{bp}
	\bp^\mu = \frac{(p+p')^\mu}{2} = \frac{-t}{2\xi Q^2 (t+Q^2)} n^\mu + \frac{Q^2}{2\xi} n'^\mu + \bp_\perp^\mu \,,
\end{equation}
where
\begin{equation}\label{p_perp^2}
	\bp_\perp^2 = M^2 - \frac{t}{4}\left( 1 - \frac{1}{\xi^2} \right)\,.
\end{equation}
Taking into account that $-\bp_\perp^{\,2} = |\bp_\perp|^2 \geq 0$, one can get the value of $t$ for which its modulus is minimal, namely $t_0$, for a fixed value of the skewness:
\begin{equation}\label{t0}
	t_0 = -\frac{4M^2\xi^2}{1-\xi^2}\qquad (|t| \geq |t_0|)\,.
\end{equation}
This way, the vector $\bp_\perp^\mu$ together with its dual $\widetilde{\bp}_\perp^\mu = \epsilon_\perp^{\mu\nu}\bp_\nu$ span the transverse plane, allowing to define transverse-polarization vectors as~\cite{Martinez-Fernandez:2025gub}
\begin{equation}\label{epsPrime}
	\eps'^\mu(\lambda) = -\frac{1}{\sqrt{2}|\bp_\perp|}\left( \bp_\perp^\mu - i\lambda\widetilde{\bp}_\perp^\mu \right)\,,\quad \lambda=\pm\,.
\end{equation}
The ``prime'' notation indicates that this vector corresponds to the outgoing photon with momentum $q'$ such that $q'\eps'(\pm)=0$. For the incoming photon, the same polarization vectors apply, this is $\eps'(\pm) = \eps(\pm)$ with $q\eps(\pm)=0$. In terms of $q$ and $q'$ we can introduce the following longitudinal-polarization vector for the incoming photon~\cite{Martinez-Fernandez:2025gub}
\begin{equation}\label{eps0}
	\eps^\mu(0) = \frac{1}{Q}\left[ \frac{2Q^2}{\scale^2}q'^\mu - q^\mu \right]\,.
\end{equation}
Here, we denoted the hard scale of the process as
\begin{equation}
	\scale^2=Q^2+t\,.
\end{equation}
Finally, the skewness, which accounts for the fractional longitudinal kick to the target, takes the form
\begin{align}
	\xi & = \frac{-\Delta n}{2\bp n} = \frac{Q^2+t}{2Q^2/x_A - Q^2 + t} \label{xi_exact} \\
	& = \frac{x_A}{2-x_A} + O\left( \frac{|t|}{Q^2} \right) \label{xi_LT} \\
	& \overset{ \mathclap{ \substack{ \rm Eq.\\(\ref{xA_vs_xB}) } } }{\approx}\ \frac{x_B}{2A-x_B} + O\left( \frac{|t|}{Q^2} \right) \label{xi_LT_with_xB}  \,.
\end{align}

\section{BH amplitude}
\label{app:BH}
\subsection{The first Bethe-Heitler amplitude, middle diagram in Fig.~\ref{fig::subprocesses}}\label{sec::amplitudes_BH}

For a configuration of beam and outgoing-photon polarizations given by $s$ and $\lambda$, respectively, the amplitude of BH reads:
\begin{equation}
\label{eq:BHAmp}
	i\M^{s\lambda}_{\rm BH} = \frac{-i e^3}{ (k-\Delta)^2  t } \eps'_\rho(-\lambda) J_\alpha \underbrace{\bar{u}(k',s) \gamma^\rho \left(\slashed{k}-\slashed{\Delta}\right) \gamma^\alpha u(k,s)}_{\mathscr{T}^{\rho\alpha}}
\end{equation}
where $e$ is the proton electric charge, $J$ is the non-elementary hadron current
\begin{equation}
	J^\mu = \langle p'| j^\mu(0) |p\rangle = 2\bar{p}^\mu F(t)\,,\quad j^\mu(0) = \sum_f e_f\bar{\quark}_f(0)\gamma^\mu\quark_f(0)\,,
\end{equation}
with $F(t)$ the electromagnetic form factor~(\ref{eq:eff}), and $\eps'(\lambda)$ is the polarization vector of the real outgoing-photon with momentum $q'$ and polarization $\lambda=\pm$, vid.~Eq.~(\ref{epsPrime}). Note that we employed the relation $(\eps'(\lambda))^* = \eps'(-\lambda)$.

In order to use the KS techniques, we need to write down the quark propagator as a superposition of lightlike vectors. Considering a massless lepton beam,
\begin{equation}
	\slashed{k} - \slashed{\Delta} = \slashed{k}'+\slashed{q}' = \sum_{h=\pm} \left[ u(k',h)\bar{u}(k',h) + u(q',h)\bar{u}(q',h) \right] \,.
\end{equation}
Then,
\begin{equation}
	\mathscr{T}^{\rho\alpha} = \sum_{h=\pm}\sum_{i=1}^2 \bar{u}(k',s) \gamma^\rho u(R_i,h)\bar{u}(R_i,h) \gamma^\alpha u(k,s) \,,
\end{equation}
where $R_i\in\{k',q'\}$ and contracting with $\eps'$ and $J$,
\begin{align}\label{eps'JT}
	\eps'_\rho(-\lambda)J_\alpha \mathscr{T}^{\rho\alpha} & = \sum_{i=1}^2 \eps'_\rho(-\lambda)J_\alpha \bar{u}(k',s) \gamma^\rho u(R_i,s)\bar{u}(R_i,s)\gamma^\alpha u(k,s) \nonumber\\
	& = -\frac{F_1\sqrt{2}}{|\bp_\perp|}\sum_{i=1}^2 \left( p_{\perp,\,\rho} + i\lambda\widetilde{p}_{\perp,\,\rho} \right)\bp_\alpha \bar{u}(k',s)\gamma^\rho u(R_i,s) \bar{u}(R_i,s)\gamma^\alpha u(k,s)\,.
\end{align}

As $\eps'$ and $J$ are given by means of $p_\perp$, $\widetilde{p}_\perp$ and $\bp$, we need to write these momenta in terms of lightlike vectors in order to apply the KS techniques. To this purpose,
\begin{align}
	p^\mu & = p^+ n'^\mu + p^- n^\mu + p_\perp^\mu \nonumber\\
	& = \underbrace{(p^+ - \alpha)n'^\mu}_{r_1^\mu} + \underbrace{(\alpha n'^\mu + p^- n^\mu + p_\perp^\mu )}_{r_2^\mu}\,,\quad r_i^2 = 0\,.
\end{align}
With $p^2=M^2$ and $r_1^2=r_2^2=0$, the parameter $\alpha$ takes the form
\begin{equation}
	\alpha = p^+ - \frac{M^2}{2p^-(nn')}\,.
\end{equation}

Consequently, the transverse component of $p$ reads
\begin{equation}
	p_\perp^\mu = r_2^\mu - p^-Q^2 q'^\mu -\alpha n'^\mu  \Rightarrow \slashed{\bp}_\perp = \sum_{j=1}^3 B_j \slashed{L}_j \,,
\end{equation}
where $B_j\in\{ 1,-p^-Q^2,-\alpha \}$ and $L_j\in\{r_2,q',n'\}$.

In a similar fashion, for the momentum of the final-state target:
\begin{align}
	p'^\mu & = p'^+ n'^\mu + p'^- n^\mu + p_\perp^\mu \nonumber\\
	& = \underbrace{(p'^+ - \alpha')n'^\mu}_{r'^\mu_1} + \underbrace{(\alpha' n'^\mu + p'^- n^\mu + p_\perp^\mu )}_{r'^\mu_2}\,,\quad r'^2_i = 0\mbox{ and }\alpha' = p'^+ - \frac{M^2}{2p'^-(nn')}\,,
\end{align}
so that
\begin{equation}
	2\bp = p+p' = r_1+r_2+r'_1+r'_2\,.
\end{equation}

Let us recover from Ref.~\cite{Deja:2023ahc} the function $g(s,\ell,a,k)$ and the current $j_\mu(s,k',k)$ given for any lightlike vectors $\ell,a,k,k'$ and helicity $s=\pm$,
\begin{align}
	g(s,\ell,a,k) & = \bar{u}(\ell,s)\slashed{a} u(k,s) \nonumber\\
	& = \delta_{s+} s(\ell,a)t(a,k) + \delta_{s-} t(\ell,a)s(a,k)\,, \label{g} \\
	j_\mu(s,k',k) & = \bar{u}(k',s)\gamma_\mu u(k,s) \nonumber\\
	& = \frac{1}{N_{k'k}}\left\{ k'_\mu (k\kappa_0) + k_\mu (k'\kappa_0) - \kappa_{0,\,\mu}(k'k) + i s \epsilon_{\mu\alpha\beta\gamma} k'^\alpha k^\beta \kappa_0^\gamma \right\}\,, \label{j}
\end{align}
where $N_{k'k} = \sqrt{(k'\kappa_0)(k\kappa_0)}$ with the auxiliary vector $\kappa_0^\mu = (1,1,0,0)$. The KS functions $s(k,k'), t(k,k')$ represent the bilinears
\begin{align}
	s(k,k') & = \bar{u}(k,+)u(k',-) = -s(k',k) \,, \label{sKS_def} \\
	t(k,k') & = \bar{u}(k,-)u(k',+) = \left[ s(k',k) \right]^* \,, \label{tKS_def}
\end{align}
that with the aid of $\kappa_0$, and as long as $k\kappa_0\neq 0$ and $k'\kappa_0\neq 0$, acquire the simple form
\begin{equation}
	s(k,k') = (k^2+ik^3)\sqrt{\frac{k'^0-k'^1}{k^0-k^1}} - (k\leftrightarrow k')\,.
\end{equation}

Making use of these expressions in Eq.~(\ref{eps'JT}), we can fully remove the Dirac spinors 
\begin{align}
	\eps'_\rho(-\lambda)J_\alpha\mathscr{T}^{\rho\alpha} & = -\frac{F(t)\sqrt{2}}{2|\bp_\perp|} \sum_{i=1}^2\sum_{\ell=1}^4 \left[ \sum_{j=1}^3 B_j g(s,k',L_j,R_i) g(s,R_i,\widetilde{r}_\ell,k) - i\lambda p_\nu\epsilon_\perp^{\nu\rho}j_\rho(s,k',R_i)g(s,R_i,\widetilde{r}_\ell,k) \right] \,,
\end{align}
and write the amplitude {\it \`a la} KS:
\begin{align}\label{iM_BH}
	i\M^{s\lambda}_{\rm BH} =&\ \frac{i e^3 F(t)}{  (k'+q')^2  t } \left(\frac{\sqrt{2}}{2|\bp_\perp|}\right) \nonumber\\
	&\ \times \sum_{i=1}^2 \sum_{\ell=1}^4 g(s,R_i,\widetilde{r}_\ell,k) \left[ \sum_{j=1}^3 B_j g(s,k',L_j,R_i) - i\lambda p_\nu\epsilon_\perp^{\nu\rho}j_\rho(s,k',R_i) \right] \,,
\end{align}
where $B_j\in\{ 1,-p^-Q^2,-\alpha \}$, $L_j\in\{r_2,q',n'\}$, $R_i\in\{k',q'\}$ and $\widetilde{r}_\ell\in\{r_1,r_2,r'_1,r'_2\}$.

\subsection{The second Bethe-Heitler amplitude, right diagram in Fig.~\ref{fig::subprocesses}}\label{sec::amplitudes_BHX}

As for the case before, for a configuration of beam and outgoing-photon polarizations given by $s$ and $\lambda$, respectively, the amplitude of the BHX subprocess reads:
\begin{equation}
	i\M^{s\lambda}_{\rm BHX} = \frac{-i e^3}{  (k-q')^2 t } J_\alpha \eps'_\rho(-\lambda) \underbrace{\bar{u}(k',s) \gamma^\alpha\left( \slashed{k} - \slashed{q}' \right)\gamma^\rho u(k,s)}_{\mathscr{T}_X^{\alpha\rho}} 
\end{equation}
Following the same steps as for BH, we find
\begin{align}
	J_\alpha\eps'_\rho(-\lambda)\mathscr{T}_X^{\alpha\rho} & = -\frac{F(t)\sqrt{2}}{2|\bp_\perp|} \sum_{i=1}^2 \sum_{\ell=1}^4 \sigma(\widetilde{R}_i) g(s,k',\widetilde{r}_\ell,\widetilde{R}_i) \left[ \sum_{j=1}^3 B_j g(s,\widetilde{R}_i,L_j,k) - i\lambda p_\nu \epsilon_\perp^{\nu\rho}j_\rho(s,\widetilde{R}_i,k) \right] \,,
\end{align}
so that the amplitude {\it \`a la} KS takes the form
\begin{align}\label{iM_BHX}
	i\M_{\rm BHX}^{s\lambda} =&\ \frac{i e^3 F(t)}{(k-q')^2 t } \left( \frac{\sqrt{2}}{2|\bp_\perp|} \right) \nonumber\\
	&\ \times \sum_{i=1}^2 \sum_{\ell=1}^4 \sigma(\widetilde{R}_i) g(s,k',\widetilde{r}_\ell,\widetilde{R}_i) \left[ \sum_{j=1}^3 B_j g(s,\widetilde{R}_i,L_j,k) - i\lambda p_\nu \epsilon_\perp^{\nu\rho}j_\rho(s,\widetilde{R}_i,k) \right] \,,
\end{align}
where, again $B_j\in\{ 1,-p^-Q^2,-\alpha \}$, $L_j\in\{r_2,q',n'\}$ and $\widetilde{r}_\ell\in\{r_1,r_2,r'_1,r'_2\}$, but $\widetilde{R}_i\in\{k,q'\}$ with the signature function $\sigma(\widetilde{R}_i) \in\{+1,-1\}$, respectively.

\section{Coefficient functions\label{app:CFs}}

In this appendix, we collect the expressions for the helicity amplitudes employed in the main text to parameterize the Compton tensor and compute cross sections and asymmetries. These formulas were published in earlier works~\cite{Braun:2022qly,Martinez-Fernandez:2025gub,Belitsky:2000jk}:
\begin{align}
    \amp^{++}(\xi,t) & = \int_{-1}^1 \frac{dx}{\xi} \Bigg\{ \Bigg[C_0(x/\xi) + \frac{\alpha_s C_F}{4\pi} C^q_{\rm NLO}(x/\xi) \nonumber\\
    & \phantom{= \int_{-1}^1 \frac{dx}{\xi} \Bigg\{} + \frac{t}{\scale^2}\left( C^{++}_{0,\,\rm HT}(x/\xi) - \xi\dxi C^{++}_{1,\,\rm HT}(x/\xi) \right) +\frac{2\xi^2\bp_\perp^2}{\scale^2}\xi^2\dxi^2 C^{++}_{1,\,\rm HT}(x/\xi) \Bigg]H^q(x,\xi,t) \nonumber\\
    & \phantom{= \int_{-1}^1 \frac{dx}{\xi} \Bigg\{} + \frac{\alpha_s T_F}{4\pi} \frac{1}{\xi} C^g_{\rm NLO}(x/\xi) H^g(x,\xi,t) \Bigg\}  \,, \\
    \amp^{+-}(\xi,t) & = \int_{-1}^1 \frac{dx}{\xi} \left\{ \frac{2\xi^2\bp_\perp^2}{\scale^2} \xi^2\dxi^2 C^{+-}_{0,\,\rm HT}(x/\xi)H^q(x,\xi,t) + \frac{\alpha_s T_F}{4\pi} \frac{2\xi^2\bp_\perp^2}{M^2} \frac{1}{\xi} C^{+-}_{\rm NLO}(x/\xi)H^g_T(x,\xi,t) \right\}  \,, \\
    \amp^{0+}(\xi,t) & = -\frac{Q \sqrt{2} \xi |\bp_\perp|}{\scale^2} \int_{-1}^1 \frac{dx}{\xi}\left(1 - \xi\dxi \right) C^{0+}_{0,\,\rm HT}(x/\xi) H^q(x,\xi,t)\,,
\end{align}

where the derivatives with respect to the skewness ($\dxi,\dxi^2$) act on everything to the right, including the GPDs. Note that the higher twists proportional to $\bp_\perp$ contain Natchman-like mass corrections and vanish as $t\to t_0$: $2\xi^2\bp_\perp^2 = (1-\xi^2)(t-t_0)/2$. The leading-twist coefficient functions ($\mu_{\rm F}$ stands for the factorization scale) read
\begin{align}
    C_0(u) & = \frac{-1}{u-1+i0} - (u\to -u) \,,\\
    C^{q/g}_{\rm NLO}(u) & = C^{q/g}_1(u) + \Ln{\frac{Q^2}{\mu_{\rm F}^2}}C^{q/g}_{\rm coll}(u) \mp (u\to -u) \quad (\mbox{$-$ for $q$, $+$ for $g$}) \,, \\
    C^q_1(u) & = \frac{1}{u+1-i0}
	\bigg[
	9 -3\frac{u+1}{u-1}\ln\left(\frac{u+1}{2}-i0\right)
	-\ln^2\left(\frac{u+1}{2}-i0 \right)
	\bigg]
	\,, \\
	C^q_{\rm coll} (u) = & 
	\frac{1}{u+1-i0}
	\bigg[
	-3-2\ln\left(\frac{u+1}{2}-i0\right) \bigg]  
	\,,\\
    C^g_1(u) =&\ \left(\sum_f e_f^2\right) \frac{1}{(u+1-i0)(u-1+i0)}
	\bigg[
    2\frac{u+3}{u-1}\ln\left(\frac{u+1}{2}-i0\right)
	-\frac{u+1}{u-1}\ln^2\left(\frac{u+1}{2}-i0 \right)
	\bigg]
	\,, \\
	C^g_{\rm coll} (u) =&\ \left(\sum_f e_f^2\right) \frac{2}{(u+1-i0)(u-1+i0)}
	\bigg[
	-\frac{u+1}{u-1}\ln\left(\frac{u+1}{2}-i0\right) \bigg]  \,, \\
    C^{+-}_{\rm NLO}(u) & = \left(\sum_f e_f^2\right) \frac{4}{(u-1+i0)(u+1-i0)}  \,,
\end{align}
while the higher-twist kernels are given by
{\allowdisplaybreaks
\begin{align}
    C^{++}_{0,\,\rm HT}(u) & = \wpbbiii(u)-\frac{\mathcal{L}(u)}{2} - (u\to -u) \,,\\
    C^{++}_{1,\,\rm HT}(u) & = \frac{\wpbbiii(u) - \mathcal{L}(u)}{2} - (u\to -u) \,,\\
    C^{+-}_{0,\,\rm HT}(u) & = u\wpbbiii(u) - (u\to -u) \,,\\
    C^{0+}_{0,\,\rm HT}(u) & = \wpbbiii(u)-(u\to -u)\,, \\
    \wpbbiii(u) & = \frac{-2}{u+1}\Ln{\frac{u-1+i0}{-2+i0}} \,,\\
    \mathcal{L}(u) & = \frac{4}{u-1}\left[\Li{2}{\frac{u+1}{2-i0}} - \Li{2}{1}\right] \,.
\end{align}
}

\end{document}